\documentclass[sigconf]{acmart}
\AtBeginDocument{%
  }

\setcopyright{acmlicensed}
\copyrightyear{2018}
\acmYear{2018}
\acmDOI{XXXXXXX.XXXXXXX}
\acmConference[Conference acronym 'XX]{Make sure to enter the correct
  conference title from your rights confirmation email}{June 03--05,
  2018}{Woodstock, NY}
\acmISBN{978-1-4503-XXXX-X/2018/06}

\usepackage{booktabs}   
\usepackage{multirow}   
\usepackage{array}      
\usepackage{arydshln}   
\usepackage{graphicx}   
\usepackage{caption}    
\usepackage{makecell}
\usepackage{tabularx}
\DeclareMathOperator*{\argmax}{arg\,max}
 \usepackage{dsfont}
\usepackage{etoolbox}
\usepackage[most]{tcolorbox}
\newtcolorbox{mybox}[1][]{
breakable,
  arc=1mm,
  boxrule=1pt,
  colback=yellow!14,
  colframe=black!80,
  fonttitle=\bfseries,
  title=#1,
  left=1mm,
  right=1mm,
  top=1mm,
  bottom=1mm
}
\usepackage{subfigure}
\begin{document}

\title{R-Log: Incentivizing Log Analysis Capability in LLMs via Reasoning-based Reinforcement Learning}

\author{Yilun Liu$^{1,2,*}$, Ziang Chen$^{1,2,*}$, Song Xu$^{2}$, Minggui He$^{2}$, Shimin Tao$^{2}$, Weibin Meng$^{2}$, Yuming Xie$^{2}$\\Tao Han$^{2}$, Chunguang Zhao$^{2}$, Jingzhou Du$^{2}$, Daimeng Wei$^{2}$, Shenglin Zhang$^{1}$, Yongqian Sun$^{1}$}\authornotemark[2]\thanks{$*$ Equal Contribution. $\dagger$ Corresponding author.} 
\affiliation{\vspace{0.05cm}\institution{$^1$ Nankai University\country{China}}
$^2$ Huawei\country{China}
}
\email{{liuyilun3, chenziang8, xusong26, heminggui, taoshimin, mengweibin3, yuming.xie, billow.han}@huawei.com}
\email{{zhaochunguang4, dujingzhou, weidaimeng}@huawei.com, {zhangsl, sunyongqian}@nankai.edu.cn}

\renewcommand{\shortauthors}{Yilun Liu, Ziang Chen, et al.}

\begin{abstract}
The growing complexity of log data in modern software systems has prompted the use of Large Language Models (LLMs) for automated log analysis. Current approaches typically rely on direct supervised fine-tuning (SFT) on log-label pairs. However, this exacerbates the domain discrepancy between general-purpose LLMs and specialized log data, causing overfitting. Furthermore, SFT's imbalanced loss computation often allows lengthy contexts to overwhelm critical, concise details in model answers, leading to hallucinations. To address these limitations, we propose R-Log, a novel reasoning-based paradigm that mirrors the structured, step-by-step analytical process of human engineers. This approach enhances generalizability by learning the underlying rules behind conclusions. We further employ Reinforcement Learning (RL) to optimize the model within a simulated O\&M environment, thereby reducing hallucinations by directly rewarding correct outcomes. R-Log is first cold-started on a curated dataset of 2k+ reasoning trajectories, guided by 13 strategies from manual O\&M practices, to establish an initial reasoning capability. This ability is then refined via RL using a joint reward function. Empirical evaluations on real-world logs show that R-Log outperforms existing methods across five log analysis tasks, particularly in unseen scenarios (by 228.05\%). We also designed R-Log-fast with 5x speedup while keeping 93\% of the efficacy.
\end{abstract}




\maketitle

\section{Introduction}\label{sec:intro}
\begin{figure}[t!]
    \centering
  \includegraphics[trim=31mm 99mm 28mm 14mm, clip, width=\linewidth]{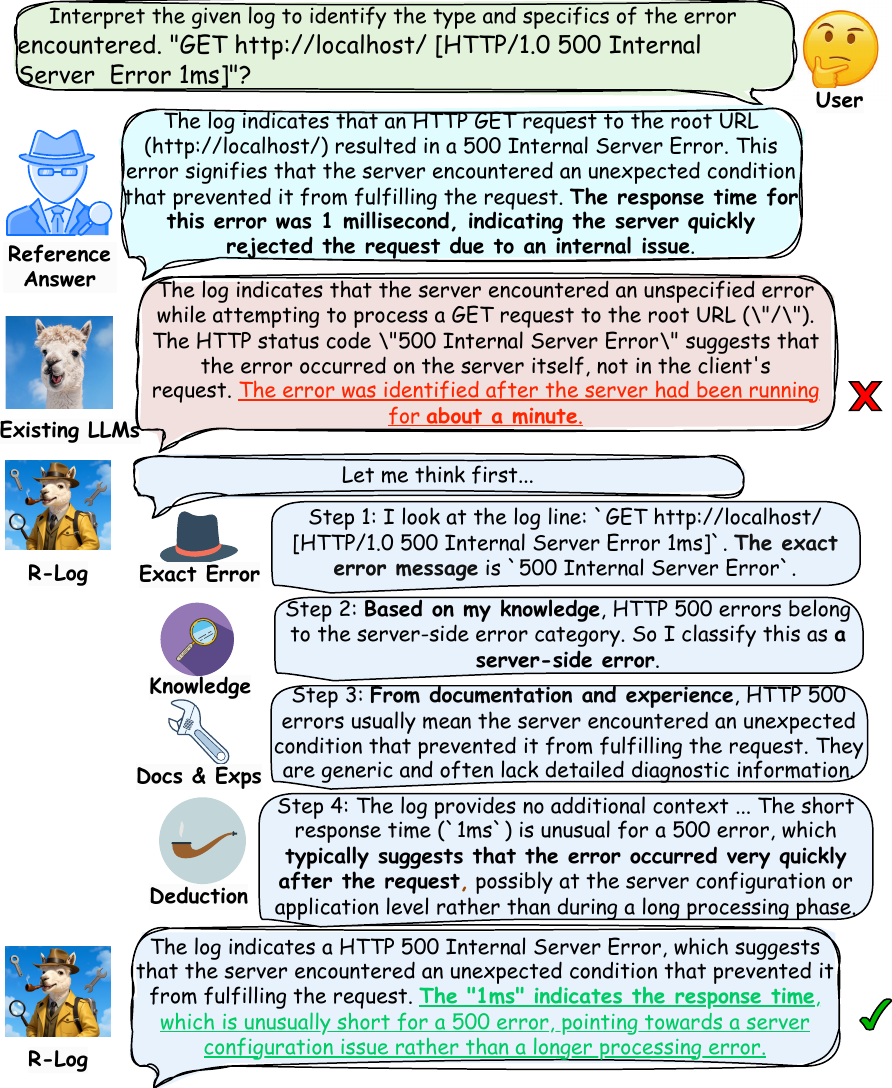}
  \caption{An real case from evaluation illustrating the reasoning-based nature of R-Log. Through human-like step-by-step reasoning, R-Log avoided the hallucinated “a minute” by existing LLMs and successfully identified the short response time “1ms” from the error log.}
  \label{fig1}
\end{figure}

Automated log analysis plays a pivotal role in modern software engineering, serving as a fundamental practice for ensuring system reliability, performance, and security. To facilitate the comprehensive understanding of diverse log events within complex software systems and aid human Operation and Maintenance (O\&M) engineers in managing the sheering volume of incidents, a variety of sub-tasks are studied in the field of log analysis. These sub-tasks range from enhancing log interpretation~\cite{liu2024logprompt} and log parsing (into variables and templates)~\cite{zhu2019tools}, to applying logs for practical problem-solving, such as anomaly detection~\cite{le2022log}, root cause analysis~\cite{chen2024automatic} and solution recommendation~\cite{liu2025loglm}.

However, the increasing complexity of software modules and the growing customization of analysis scenarios pose new challenges: training and deploying massive specialized models for each scenario, task or domain can lead to significant cost~\cite{The_State_of_LLM_Operations_or_LLMOps,liu2025loglm}. In practice, the performances of these specialized models are further hampered by insufficient historical logs, as online environments may continuously generate new log patterns unseen during training~\cite{liu2024interpretable,liu2024logprompt}. To address this, Liu~\emph{et al.} proposed LogLM~\cite{liu2025loglm}, fine-tuning a single large language model (LLM) for multiple sub-tasks of log analysis, given LLMs' strong capability acquired in pre-training~\cite{dubey2024llama3}. 

Nevertheless, LLMs are typically trained on generic plain text corpora (\emph{i.e.}, natural language), which unavoidably introduces distribution discrepancy~\cite{smithunderstanding, ji2025superlog} when being directly fine-tuned on domain-specific and highly structured log texts (\emph{e.g.}, IP addresses, status codes and file paths, \emph{etc.}). For example, LogLM relies on direct answer fitting via Supervised Fine-Tuning (SFT) using log-label pairs without learning any rules behind the analytical conclusions. In this paper, we hypothesize that, incentivizing capability in general-purpose LLMs for specialized, complex domains like log analysis may require a paradigm shift: \textbf{from directly fitting answers (X→Y) to learning the reasoning trajectories of human experts in natural language to derive answers through reasoning (X→(R,Y))}. This shift is inspired by observations on real-world software development and O\&M practices, where human engineers decompose complex problems from unfamiliar domains into sub-steps and follow a trajectory (\emph{e.g.}, procedures outlined in O\&M manuals~\cite{guideline1992root}) to reason toward final conclusions~\cite{jeffries2013processes}. Secondly, with the increasing of task contexts and scenarios complexity, direct SFT on longer and more heterogeneous sequences (\emph{e.g.}, multi-task training) often struggles to converge to the optimal~\cite{chusft}, suggesting the need for more sophisticated learning algorithms tailored to complex scenarios (\emph{e.g.}, Reinforcement Learning (RL)).

To this end, we propose R-Log, a novel paradigm to incentivize log analysis capabilities in LLMs by driving the model to reason (\emph{i.e.}, output thinking trajectories as shown in Fig.~\ref{fig1}) first before getting the answer through RL. The key innovations of R-Log are:

\textbf{(1) From X→Y to X→(R,Y).} Existing methods directly train LLMs on log (X)-label (Y) pairs, expecting the model to spontaneously learn intrinsic patterns for solving log analysis problems. However, the distribution discrepancy between general texts and logs makes LLMs difficult to learn the intrinsic patterns, potentially leading to overfitting on historical logs (\emph{i.e.}, learning by rote~\cite{huang2025logrules}). Consequently, the X→Y paradigm may result in limited generalizability to handle unseen scenarios (\emph{e.g.}, performance of LogLM degrades sharply for a untrained new task in Section~\ref{sec:RQ4}). In contrast, R-Log’s learning objective encompasses both the answer and an explicit reasoning process aligned with human practices (\emph{i.e.}, X→(R,Y)), facilitating the learning of universal rules behind the answers (\emph{e.g.}, the steps in Fig.~\ref{fig1}), thereby enhancing generalization (\emph{i.e.}, outperforming LogLM in the unseen task by 228.05\%). Additionally, the model's explicit reasoning trajectory enhances its transparency and trustworthiness, forming the core feature of its deployment at Huawei (discussed in Section~\ref{sec:in practice}).

\textbf{(2) From SFT to RL.} With the increasing complexity of log analysis scenarios, handling complex tasks with flexible instructions and heterogeneous answers has become a preferred requirement for LLMs~\cite{liu2025loglm}. However, SFT inherently optimizes in a token-by-token “memorization” manner~\cite{chusft}, where losses are evenly distributed across tokens (\emph{i.e.}, a word-wise comparison with reference answers), making convergence difficult for long and complex sequences~\cite{helm2025token}. 

For instance, in Fig.~\ref{fig1}, a user asks to interpret the error log with natural language (a popular scenario in recent studies~\cite{liu2024interpretable,liu2025loglm}), and the reference answer is complex beyond a short label (\emph{e.g.}, "normal" or "abnormal"). Since the computation of loss in SFT is evenly distributed across the long reference answer, some minor key point like the “1ms” will be overwhelmed by other information, leading to hallucinations (\emph{i.e.}, “1ms” was identified as “a minute” by existing LLMs in Fig.~\ref{fig1}). Another case is when fine-tuning the model with reasoning trajectories, the longer reasoning part $R$ receives more signals than the shorter $Y$, potentially preventing the model from producing correct conclusions (further discussed in Section~\ref{sec:RQ2})\footnote{A straightforward solution may be to balance the loss with different weights, but it is hardly feasible in practice to precisely label different parts and assign proper weights.}. R-Log overcomes this by modeling log analysis as an O\&M agent interacting with a heterogeneous environment that simulates real-world O\&M practices. The agent takes actions on system-triggered log incidents that require diverse analysis skills (\emph{e.g.}, log parsing, or root cause analysis), receives rewards based on the correctness of its analysis, and updates parameters using RL algorithms. Therefore, it better aligns with human practices and prioritizes correct conclusions over mere token matching, which reduces hallucinations and increases generalization in complex scenarios as the improvement by RL stage in Fig.~\ref{fig_ablation}(a) and Table~\ref{tab:unseen task}.

To achieve this, we first collected expertise from human log analysts and summarized it into 13 reasoning templates that reflects thinking patterns of professional engineers in various log analysis tasks. We then instantiated these reasoning templates with real-world logs and constructed a dataset of 2k+ samples with X→(R,Y) pairs, named the Log Reasoning Dataset. R-Log’s training is two-staged: in the first stage, we perform SFT on the base model using Log Reasoning Dataset as a cold-start, enabling the LLM to imitate human expert reasoning strategies; in the second stage, we employ RL on the cold-started model to further calibrate its reasoning paradigms. Evaluation results indicate that R-Log outperforms specialized models and general-purpose LLMs across five log analysis tasks, particularly in handling unseen tasks. Our contributions are:
\begin{itemize}
\item We propose a novel reasoning-based log analysis paradigm that enables LLMs to learn the reasoning strategies of human experts rather than merely fitting labels, enhancing both generalization and interpretability.
\item We design a novel RL-based training algorithm for log analysis, allowing the model to adapt to more complex scenarios and contexts.
\item We open-source the Log Reasoning Dataset\footnote{Code, dataset and reasoning templates available at \url{https://github.com/lunyiliu/R-Log}} containing 2k+ real-world reasoning trajectories in log analysis as well as the 13 curated typical reasoning strategies from manual practices, facilitating future research.
\end{itemize}

\section{Related Work}
\subsection{Reasoning-based LLMs}

Since the pioneering advancement of DeepSeek-R1~\cite{deepseek2025r1} on solving math problems, the reasoning-based paradigm (also called inference-time scaling) has become a promising alternative for LLMs on solving complex, domain-specific and expertise-required problems. Reasoning-based LLMs are characterized by its scaled reasoning steps before outputting final answer during inference time, thereby are particularly suitable for domains that naturally require a thinking procedure to reach the conclusion. For example, in addition to mathematics, the advantage of reasoning-based LLMs has been further verified on domains such as code generation~\cite{li2025s}, legal field~\cite{chu2025domaino1s}, financial reasoning~\cite{qian2025fino1} and machine translation~\cite{he2025r1}.

Given the complexity and expertise required to perform the tasks of log analysis, as well as the pervasive existence of reasoning procedures as recommended by manuals~\cite{guideline1992root, kent2006nist} in real-world software O\&M practices, it is motivated for reasoning-based paradigm to be applied in it. And R-Log advances by demonstrating the advantage of reasoning-based LLMs in software log analysis, through its carefully designed training paradigm tailored for log analysis.

\subsection{Reinforcement Learning}~\label{sec:RL intro}

With the increasing complexity in applied scenarios of AI systems, the theory of RL develops to overcome the learning goals that are infeasible for traditional gradient-based algorithms (\emph{e.g.}, non-differentiable)~\cite{wang2022deep}. By simulating the real-world learning environment, an agent takes actions according a certain probability distribution (called policy) and receives reward signals from the environment, iteratively converging to an optimized policy for the specific environment~\cite{botvinick2020deep}. Recently, RL techniques have achieved significant success in building powerful LLMs, aligning the model with human expectations through the RLHF (Reinforcement Learning with Human Feedback) algorithm. Compared with RLHF which requires training a critic model to simulate human feedbacks, another popular alternative is GRPO (Group Relative Policy Optimization)~\cite{shao2024deepseekmath}, a simplified RL algorithm adopted by DeepSeek-R1 without training a critic model, which uses relative advantages among a group of sampled answers to define the reward signals. By utilizing GRPO, recent studies have explored performing software failure localization and change management~\cite{zhang2025thinkfl,sun2025enhancing}. 

Compared with existing studies, R-Log firstly applied RL into the general field of log analysis. Beyond a specific sub-task (\emph{e.g.}, log anomaly detection), we designed a joint reward function to measure actions in a heterogeneous O\&M environment with various analysis tasks, enabling a more generalized policy for log analysis.

\subsection{Log Analysis}
\subsubsection{Task-specific Approaches}~\label{sec:log task intro}

To handle a specific challenge or facilitate a specific analysis scenario, new sub-tasks in log analysis and specialized approaches for the sub-tasks continuously emerges. Typical sub-tasks are:

\textit{(1) Log Parsing}, which aims at parsing raw logs into templates and variables, thereby reducing its sheering volumes and facilitates further analysis. A log template only retains static parts in a log and replaces dynamic variables with a symbol of <*>. Log parsing approaches can be divided into coarse-level methods, which focus on extracting common parts as templates in raw logs~\cite{fu2009execution,du2016spell,makanju2009clustering,zhang2017syslog,he2017drain,meng2020summarizing}, and fine-level methods, which focus on identifying variables within raw logs~\cite{meng2020logparse,tao2022logstamp,huo2023semparser,li2023did}.

\textit{(2) Log Anomaly Detection}, which classifies anomalous events from normal behaviors within system logs. Traditional approaches require massive historical logs as training samples to model the anomaly patterns within systems~\cite{du2017deeplog,meng2019loganomaly,zhang2019robust}, thereby can hardly handle the frequently updated online environment. In contrast, semantic-based approaches identifies anomalous patterns based on understanding semantics within log templates~\cite{liu2024interpretable,pan2024raglog,qi2023loggpt}, thereby can achieve better results with a small amount of training samples.

\textit{(3) Log Interpretation}, which aims at aiding human engineers in interpreting logs by describing the logs in natural language. Liu~\emph{et al.} proposed a systematic criteria for evaluating the performance of log interpretation and tested LLMs on this sub-task using specifically designed prompts~\cite{liu2024interpretable,liu2024logprompt}.

\textit{(4) Root Cause Analysis}, which predicts the root causes of system events recorded by logs. Chen~\emph{et al.}~\cite{chen2024automatic} implemented an LLM-empowered system that predicts cloud incidents' root cause type.

\textit{(5) Solution Recommendation}, which provides mitigating solutions to crashes indicated by system logs. Ahmed~\emph{et al.}~\cite{ahmed2023recommending} firstly explored this sub-task by fine-tuning LLMs to handle cloud incidents and recommend mitigation steps for engineers.

\textit{(6) Log Variable Classification}, which recognizes the categories of dynamic variables within logs to aid downstream analysis. Compared with log parsing which merely identifies appearance of variables, this sub-task requires a more comprehensive understanding to logging patterns within the system and predicts the specific type of variables (\emph{e.g.}, an string of numbers can either be an object amount or a status code)~\cite{li2023did,huo2023semparser}.

\subsubsection{Unified Approaches}

With the development of LLMs, the boundary between sub-tasks in log analysis has been blurred~\cite{zhu2021unilog}. By modeling log-label pairs in different sub-tasks into a unified format of instruction-response, LogLM achieved utilizing a single LLM to perform various log analysis tasks~\cite{liu2025loglm}. Ji~\emph{et al.} utilize continuous pre-training on interpretable knowledge to transform a general-purpose LLM into a specialized model tailored for log analysis~\cite{ji2025superlog}. Gou~\emph{et al.}~\cite{guoowl} proposed OWL, an LLM trained for question-and-answering (QA) tasks in IT operation.

However, existing approaches fit the LLM with answers only, limiting its potential in understanding reasons behind analysis answers. In contrast, R-Log is trained on human-aligned reasoning trajectories and is further enhanced through RL, thereby better incentivizing log analysis capabilities within LLMs.    

\section{Methodology}
\begin{figure*}[t!]
    \centering
  \includegraphics[width=\linewidth]{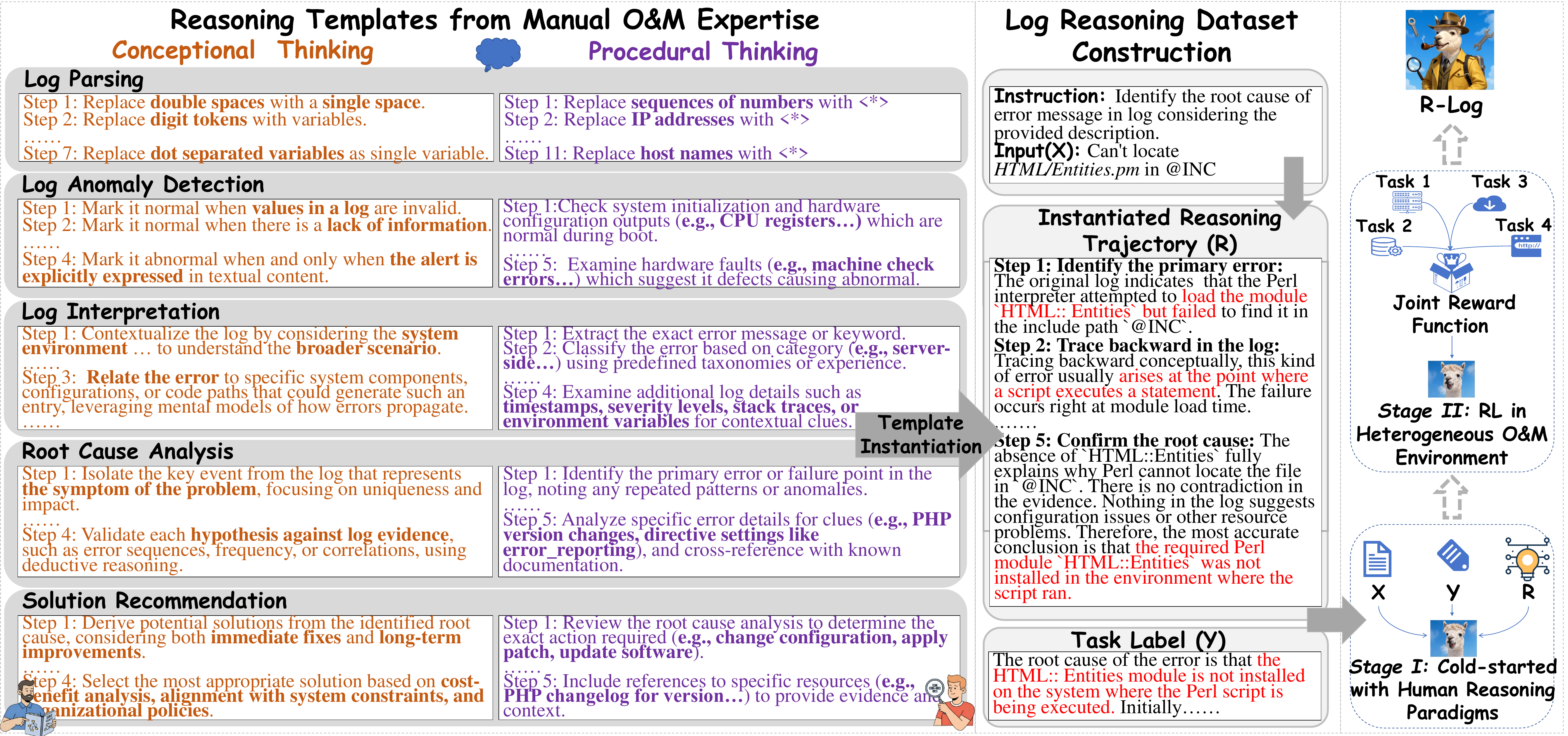}
  \caption{Illustration on the construction of the human-aligned Log Reasoning Dataset and two-staged training of R-Log. Highlighted parts reflects natures of “conceptional” or “procedural” in thinking strategies, and the real-world logs in instantiation.}
  \label{fig_main}
\end{figure*}
An overview of R-log is presented in Fig.~\ref{fig_main}. We first studied typical reasoning strategies human O\&M engineers employs when conducting log analysis in practice, and summarized into 13 task-specific reasoning templates tailored for various log analysis sub-tasks. Then, we instantiated these templates with real-world logs into reasoning trajectories (R) from the log to be analyzed (X) to the conclusion (Y), handling five specific log analysis sub-tasks. Using this dataset containing X→(R,Y) samples, we trained R-Log with a two-stage strategy. In the first stage, by directly SFT the foundation model using X→(R,Y) samples, the model is cold-started with the pre-constructed reasoning paradigms imitating human experts. In the second stage, we designed a joint reward function to guide the further optimization of reasoning paradigms in a heterogeneous O\&M environment with diverse log analysis scenarios.  

\subsection{Reasoning Strategies in Manual O\&M Expertise}

As revealed by Jefferies~\emph{et al.}~\cite{jeffries2013processes} in a psychological study on human software engineers, senior engineers tend to “decompose a problem more richly into minimally interacting parts,” suggesting the importance of reasoning strategies in software engineering. In order to built a comprehensive and human-aligned reasoning ability for LLMs, we started by investigating reasoning strategies of human O\&M engineers in log analysis, from both the angle of cognitive psychology and task-specific practice guidelines.

\subsubsection{Conceptional Thinking \emph{v.s.} Procedural Thinking}

By studying individual cognitive differences of software engineers, Bill~\cite{curtis1984fifteen} concluded two primary ways to build reasoning strategies for engineers: (1) by abstracting from development experience and (2) by learning specific rules from training programs. This taxonomy, aligning with modern cognitive psychology~\cite{tall2001symbols}, reveals two distinct reasoning patterns adopted by software engineers: conceptional thinking and procedural thinking. In the context of software log analysis, conceptional thinking style seeks to abstract patterns and strategies from experiences to guide the log analysis tasks (\emph{e.g.}, judging anomaly by overall semantics in logs), while procedural thinkers tend to follow a checklist-like reasoning trace focusing on examining specific attributes (\emph{e.g.}, checking specific status codes in logs to determine failure). These distinct thinking styles reflect the diversity of human reasoning strategies, and guide through our dataset curation.

\subsubsection{Task-specific Reasoning Strategies}

To ensure a comprehensive coverage of analysis scenarios, we investigated human reasoning strategies in five typical log analysis sub-tasks (Section~\ref{sec:log task intro} introduced these sub-tasks), encompassing both conceptional and procedural thinking styles. The sources of these strategies include manually-extracted rules for performing sub-tasks~\cite{huang2025logrules, liu2024interpretable}, software O\&M manuals and guidelines~\cite{kent2006nist,guideline1992root} and interviews with O\&M practitioners. 

\paragraph{Reasoning Strategies for Log Parsing and Anomaly Detection} 

These two sub-tasks have been widely investigated by community. Liu~\emph{et al.}~\cite{liu2024interpretable} proposed a series of high-level steps (\emph{e.g.}, judging by semantics like “textual alert content”) to perform anomaly detection, which we utilize as a conceptional reasoning strategy for anomaly detection. Huang~\emph{et al.}~\cite{huang2025logrules} extracted procedural and conceptional rules for both log parsing and anomaly detection. Specifically, the procedural reasoning strategy for log parsing focuses on examining specific variable types such as “replacing IP addresses with...” or “replacing time indicators with...”; while the conceptional strategy abstract several patterns for being recognized as variables, such as “digit tokens” and “path-like token”. These prior knowledge from practice can serve as a reasoning paradigm for log analysis. 

\paragraph{Reasoning Strategies for Log Interpretation, Root Cause Analysis and Solution Recommendation (IRS)} 

The IRS sub-tasks are not as well-studied as the two tasks above in automated log analysis. However, interpreting log content, finding root cause and employing solutions are essential skills for O\&M engineers in practice. Therefore, we seek wisdom from existing software O\&M manuals which provide strategic guidelines for engineers. An official O\&M guidance document~\cite{guideline1992root} stated recommended steps for engineers to locate root causes, from recognizing key problems in logs, determining direct cause, to tracing back to the root cause. Kent~\emph{et al.}~\cite{kent2006nist}, in a log management guidance, suggested thinking strategies for proposing practical solutions for logged errors. The strategy involves self-reflection, proposing several possible corrective actions and assessing them to determine a optimized solution. 

In addition, we also make interviews with a small group of O\&M practitioners from Huawei to share their reasoning strategies for the IRS scenarios. Each interview usually lasts 20 to 30 minutes, the interviewee is queried to share their experiences in solving triggered system problems, from interpreting error logs, locating root causes, to taking mitigating actions. From their shared cases and experiences, both conceptional and procedural thinking strategies are summarized by authors. For instance, the conceptional thinking strategy for the sub-task of log interpretation involves contextualizing with broader scenario, emphasizing general implications of logs and relating the logs with system environments before generating the final interpretations; while the procedural strategy firstly searches for existence of specific error message, classifies the error categories and examining attributes such as timestamp or variables.

\subsection{Construction of Log Reasoning Dataset}

\begin{table}[tbp]
\caption{Statistics on Reasoning Templates and Log Reasoning Dataset.}
\centering
\resizebox{\linewidth}{!} {%
\begin{tabular}{c@{\hskip 0.001in}c@{\hskip 0.001in}c@{\hskip 0.001in}c@{\hskip 0.001in}c@{\hskip 0.001in}c}
\toprule
\multicolumn{1}{l}{\textbf{Tasks$^{\mathrm{a}}$}} & \begin{tabular}[c]{@{}l@{}} \multicolumn{1}{c}{\textbf{\# Reasoning}} \\ \multicolumn{1}{c}{\textbf{Templates}} \end{tabular}& \begin{tabular}[c]{@{}l@{}} \multicolumn{1}{c}{\textbf{Template}} \\ \multicolumn{1}{c}{\textbf{Source}} \end{tabular} & \begin{tabular}[c]{@{}l@{}} \multicolumn{1}{c}{\textbf{\# X→(R,Y)}} \\ \multicolumn{1}{c}{\textbf{Samples}} \end{tabular} & \begin{tabular}[c]{@{}l@{}} \multicolumn{1}{c}{\textbf{Avg. Len.}} \\ \multicolumn{1}{c}{\textbf{of R}} \end{tabular} & \multicolumn{1}{l}{\textbf{Domain}}\\
\midrule
\multicolumn{1}{l}{\multirow{7}{*}{\textbf{L.P.}}} & \multirow{7}{*}{2} & \multirow{7}{*}{\begin{tabular}[c]{@{}l@{}} \multicolumn{1}{c}{Extracted} \\ \multicolumn{1}{c}{Rules} \end{tabular}} &  200 & 128.97 & HDFS\\
 &  & & 200 & 105.70 & Hadoop\\
 &  & & 200 & 83.14 & Zookeeper\\
 &  & & 200 & 61.66 & BGL\\
 &  & & 200 & 80.41 & HPC\\
 &  & & 200 & 114.80 & Linux\\
 &  & & 200 & 124.64 & Proxifier\\
 \midrule
\multicolumn{1}{l}{\multirow{2}{*}{\textbf{A.D.}}} & \multirow{2}{*}{3} & \multirow{2}{*}{\begin{tabular}[c]{@{}l@{}} \multicolumn{1}{c}{Extracted} \\ \multicolumn{1}{c}{Rules} \end{tabular}} &  194 & 98.28 & BGL\\
 &  & &  138 & 102.98 & Spirit\\
 \midrule
\multicolumn{1}{l}{\textbf{L.I.}}& 2 & \multirow{3}{*}{\begin{tabular}[c]{@{}l@{}} \multicolumn{1}{c}{Manuals \&} \\ \multicolumn{1}{c}{Interviews} \end{tabular}} &  300 & 257.47 & \multirow{3}{*}{Apache}\\
\multicolumn{1}{l}{\textbf{R.C.}}& 3 & &  300 & 320.97 & \\
\multicolumn{1}{l}{\textbf{S.R.}}& 3 & &  300 & 344.76 & \\
\hdashline
\noalign{\vskip 4pt}
\multicolumn{1}{l}{All}& 13 & &  2632 & 171.01 & -\\
\bottomrule
\multicolumn{6}{l}{$^{\mathrm{a}}$ \textbf{L.P.}, \textbf{A.D.}, \textbf{L.I.}, \textbf{R.C.}, \textbf{S.R.} represent the sub-tasks of log parsing, anomaly}\\
\multicolumn{6}{l}{~~\,detection, log interpretation, root cause analysis, solution recommendation.}\\
\end{tabular}
}
\label{tab:Log Reasoning Dataset}
\end{table}

\subsubsection{Human Reasoning Templates for Log Analysis}

Upon acquiring the reasoning strategies, we crafted them into reasoning templates with a unified format of “Step 1:...” to “Step $n$:..” until reaching the analysis goals. As shown in Fig.~\ref{fig_main}, each sub-task is equipped with at least one conceptual and one procedural thinking strategy, ensuring the diversity of our reasoning templates. Statistics and sources of these templates are shown in Table~\ref{tab:Log Reasoning Dataset}. For a sub-task $S_i$, we denote the template set for this sub-task as $T_i$, $i\in [1,5]$. These reasoning templates reflect how human O\&M engineers reasons in different log analysis scenarios and guide through our curation of the instantiated reasoning trajectories in Log Reasoning Dataset.

\subsubsection{Instantiating Templates to Reasoning Trajectories for Real-world Logs}

To facilitate the learning of human thinking patterns for LLMs, we instantiated these curated templates into diverse reasoning trajectories handling real-world logs. In other words, under a specific analysis task, the specific reasoning process (R) to reach task-expected outputs (Y) given real-world logs (X), is generated following the reasoning steps in templates. To ensure a comprehensive coverage of various analysis scenarios, we use the open-source training dataset from LogLM~\cite{liu2025loglm} as the initial data source. This dataset contains a total of 2600+ (X,Y) samples, encompassing five distinct log analysis sub-tasks, with real-world logs from diverse domains such as supercomputers, distributed systems, operating systems, and software applications. For every (X,Y) sample of sub-task $S_i$ in the source dataset, the goal of the instantiation is to generate the reasoning trajectory R from X to Y, under the guidance of a random template $t\in T_i$.

Following recent studies~\cite{he2025r1,liu2024you}, we automation this process using an advanced LLM by a role-playing prompt~\cite{kong2023better}, which assumes the LLM to be an O\&M engineer who needs to analyze a given log to reach the expected conclusion and asks the LLM to output the inner monologue of the engineer guided by the reasoning template. The prompts are slightly fine-tuned to suit different sub-tasks, and the version for anomaly detection is as follows:
\begin{mybox}
Assume that you are a DevOps engineer with extensive experience in log analysis and a strong ability to detect anomalies in logs. Now you have both the unstructured log and its labeled status (normal or abnormal), Your task is to construct the entire analysis process from Original Log to Log Label into an inner monologue, based on the following Reasoning Guidance: \textcolor{red}{\{Reasoning Template\}}. You must strictly follow Reasoning Guidance step by step without skipping steps and output the chain-of-thought trajectory from Original Log to Log Label. Note that the monologue should purely be starting from the Original Log without leaking any information in the labeled status, since it reflects the step-by-step internal mental activity of the engineer who doesn’t know the answer at first. The reference to the guidance should be specific combining the input log, avoiding using vague phrases like 'according to Step *'. Make sure every reasoning step in the monologue exists in the actual given log characteristics and do not propose anything outside it. Original Log: \textcolor{red}{\{Log X\}}; Log Label: \textcolor{red}{\{Label Y\}}.
\end{mybox}

Statistics of the final Log Reasoning Dataset is shown in Table~\ref{tab:Log Reasoning Dataset}. Guided by the curated 13 human-aligned reasoning templates, 2600+ distinct reasoning trajectories R are generated from (X,Y) pairs, with an average length of 171.01 words. Notably, analysis output Y for the three IRS tasks is generally longer than the sub-task of log parsing and anomaly detection (\emph{i.e.}, chunks of analytic words \emph{v.s.} a parsed template or a conclusion of normal/abnormal), thereby leading to a significantly longer reasoning trajectories R in average.

\subsection{Training of R-Log}

\subsubsection{Stage I: Cold Start on Log Reasoning Dataset}

Recent studies on reasoning-based LLMs reveal the importance of initial policy at the beginning of RL~\cite{deepseek2025r1,he2025r1}, which not only facilitates the convergence of policy, but also pose a significant impact on the final learnt policy through RL. The human prior knowledge in our Log Reasoning Dataset can provide a sound paradigm for the model to imitate during the cold-start phase, where the model acquires its initial reasoning abilities on log analysis tasks and serves as a sound initial policy for the RL stage. 

To achieve this, we simply perform a token-by-token SFT on the foundation model $\theta$, using the Log Reasoning Dataset, denoted by $D$. For the convenience of reward computing in the RL stage, $R$ and $Y$ are wrapped by two pairs of special tokens into “<think>$R$</think>” and “<answer>$Y$</answer>” in the desired outputs to control the format. The learning goal of the cold-start phase becomes:
\begin{equation}\label{eq1}
   \theta_c = \argmax_{\theta} \sum\limits_{(X,Y,R) \in D} \log P(\,  [R ; Y] \,|\, [I; X], \theta),
\end{equation}
where $I$ is a short task-specific descriptional instruction accompanying every (X,Y) sample in the source dataset (\emph{e.g.}, “Find the root cause of errors in the following log.”) and $[*;*]$ means concatenation of two strings.

\subsubsection{Stage II: RL on Joint Rewards in Heterogeneous O\&M Environment}

In the second stage, we further optimize the cold-started policy $\theta_c$ through RL to enhance its reasoning capabilities in a heterogeneous O\&M environment. By reusing (X,Y) samples in the Log Reasoning Dataset, this environment is characterized by its diversity, simulating system-side events encompassing five distinct log analysis sub-tasks (as denoted by $S_i$ for $i \in [1,5]$) and logs sourced from various domains. The agent (\emph{i.e.}, the LLM with an initial policy $\theta_c$) interacts with the environment by taking an action (\emph{i.e.}, generating a reasoning trajectory $R'$ and final answer $Y'$) based on the current state (the input log $X$ and its accompanying instruction $I$ simulating system-triggered events). The environment then provides a reward signal $r$ to guide the policy optimization. For the specific parameter updating strategy, we employ the GRPO algorithm (as introduced in Section~\ref{sec:RL intro}), which demonstrates effective in various complex reasoning scenarios~\cite{he2025r1,deepseek2025r1,zhang2025thinkfl}, by employing group-based advantage estimation. The GRPO algorithm first samples multiple answers from the policy for the same input, compute reward $r$ for each answer, and encourages the answers with a higher relative reward among the group.

The joint reward function $r$ is designed as follows to assess the quality of the agent's generated response to various events in the simulated environment: 
\begin{equation}\label{eq2}
r =
\resizebox{0.93\linewidth}{!}{%
$\displaystyle
\begin{cases}
-20 \cdot \omega_f & \text{if } \delta_f=0 \\
    \begin{aligned}
        & \omega_f+\text{F1-score}(Y, Y')\cdot \delta_v + \\
        & \quad (1-\delta_v)(1 - \dfrac{\text{EditDistance}(Y, Y')}{\max(|Y|, |Y'|)}) \\
    \end{aligned}  & \text{if } i = 1 \\
\omega_f+  \mathds{1}(Y' = Y) & \text{if } i = 2 \\
\omega_f+ \dfrac{ \text{BLEU}(Y, Y') + \sum_{n\in{1,2,L}} \text{ROUGE-}n(Y, Y') }{400} & \text{if } i \in {3,4,5}.
\end{cases}
$
}
\end{equation}

The first term is a format checking that penalizes structural errors. If the response strictly adheres to the required structure, containing both a reasoning part enclosed in “<think>...</think>” and a non-empty answer part enclosed in “<answer>...</answer>”, $\delta_f$ is $1$ and the model will receive a small reward. If the format checking fails, $\delta_f$ becomes $0$ and a severe penalty is applied. $\omega_f$ is a hyperparameter controlling weights of the format reward. Following He~\emph{et al.}~\cite{he2025r1}, we set $\omega_f$ to be 0.1, a small value to avoid distraction from answer quality. All answer rewards are normalized to $[0,1]$ to ensure consistent optimization across tasks, making the overall range of $r$ to be $[-2,1.1]$.

Upon passing format checking, $r$ is determined by evaluating the extracted answer $Y'$ based on the specific sub-task $S_i$:

\begin{table*}[tbp]
\caption{Comparing R-Log with Existing Methods on the Sub-task of Log Parsing.}
\centering
\resizebox{0.95\linewidth}{!} {%
\setlength{\tabcolsep}{2pt}
\begin{tabular}{l@{\hskip 0.1in}c@{\hskip 0.05in}c@{\hskip 0.1in}c@{\hskip 0.05in}c@{\hskip 0.1in}c@{\hskip 0.05in}c@{\hskip 0.1in}c@{\hskip 0.05in}c@{\hskip 0.1in}c@{\hskip 0.05in}c@{\hskip 0.1in}c@{\hskip 0.05in}c@{\hskip 0.1in}c@{\hskip 0.05in}c@{\hskip 0.1in}||c@{\hskip 0.1in}c@{\hskip 0.05in}c}
\toprule
\multirow{2}{*}{{\textbf{Methods}}} & \multicolumn{2}{>{\hspace{-1em}\centering}c}{\textbf{HDFS}} & \multicolumn{2}{>{\hspace{-1em}\centering}c}{\textbf{Hadoop}} & \multicolumn{2}{>{\hspace{-1em}\centering}c}{\textbf{Zookeeper}} & \multicolumn{2}{>{\hspace{-1em}\centering}c}{\textbf{BGL}} & \multicolumn{2}{>{\hspace{-1em}\centering}c}{\textbf{HPC}} & \multicolumn{2}{>{\hspace{-1em}\centering}c}{\textbf{Linux}} & \multicolumn{2}{>{\hspace{-1em}\centering}c}{\textbf{Proxifier}} & \multicolumn{2}{c}{\textbf{Avg.}} \\ \cmidrule(l{-0.3em}r{0.8em}){2-3} \cmidrule(l{-0.3em}r{0.8em}){4-5} \cmidrule(l{-0.3em}r{0.8em}){6-7} \cmidrule(l{-0.3em}r{0.8em}){8-9} \cmidrule(l{-0.3em}r{0.8em}){10-11} \cmidrule(l{-0.3em}r{0.8em}){12-13} \cmidrule(l{-0.3em}r{0.8em}){14-15} \cmidrule(lr){16-17}
& \textbf{RI$^{\mathrm{a}}$} & \textbf{F1} & \textbf{RI} & \textbf{F1} & \textbf{RI} & \textbf{F1} & \textbf{RI} & \textbf{F1} & \textbf{RI} & \textbf{F1} & \textbf{RI} & \textbf{F1} & \textbf{RI} & \textbf{F1} & \textbf{RI} & \textbf{F1}\\
\midrule
\addlinespace
Drain~\cite{he2017drain} & 0.914 & 0.389 & 0.647 & 0.068 & 0.787 & 0.225 & 0.822 & 0.397 & 0.119 & 0.002 & 0.695 & 0.225 & 0.822 & 0.500 & 0.687 & 0.258 \\
LogStamp~\cite{tao2022logstamp} & 0.954 & 0.523 & 0.927 & 0.594 & 0.992 & 0.275 & 0.984 & 0.818 & 0.949 & 0.434 & 0.760 & 0.658 & 0.811 & 0.438 & 0.911 & 0.534 \\
LogLM-7B~\cite{liu2025loglm} & 0.878 & 0.815 & 0.857 & 0.671 & 0.910 & 0.709 & 0.859 & 0.545 & 0.887 & 0.508 & 0.796 & 0.866 & 0.568 & 0.799 & 0.822 & 0.702 \\
Qwen2.5-7B-Instruct\cite{Yang2024Qwen25TR} & 0.923 & 0.832 & 0.896 & 0.583 & 0.946 & 0.767 & 0.914 & 0.482 & 0.938 & 0.682 & 0.810 & 0.768 & 0.683 & 0.833 & 0.870 & 0.707 \\
Qwen2.5-7B-SFT & 0.900 & 0.706 & 0.902 & 0.768 & 0.936 & 0.809 & 0.988 & 0.839 & 0.949 & 0.519 & 0.773 & 0.723 & 0.559 & 0.759 & 0.861 & 0.732 \\
LogPrompt~\cite{liu2024logprompt} & 0.890 & 0.863 & 0.879 & 0.763 & 0.948 & 0.889 & 0.964 & 0.865 & 0.934 & 0.759 & 0.758 & 0.766 & 0.567 & 0.653 & 0.849 & 0.794 \\
LILAC~\cite{jiang2024lilac} & \textbf{1.000} & 0.741 & 0.999 & 0.730 & 0.999 & 0.852 & \textbf{0.999} & 0.834 & \textbf{0.998} & 0.875 & 0.899 & 0.786 & 0.696 & 0.928 & 0.942 & 0.821 \\
PreLog~\cite{le2024prelog} & 0.993 & 0.705 & \textbf{1.000} & 0.815 & \textbf{1.000} & 0.979 & \textbf{0.999} & 0.934 & \textbf{0.998} & 0.873 & 0.886 & 0.798 & 0.928 & 0.859 & 0.972 & 0.852 \\
SuperLog-7B~\cite{ji2025superlog} & 0.978 & 0.977 & 0.982 & 0.942 & 0.998 & 0.815 & 0.976 & 0.700 & 0.974 & 0.727 & \textbf{0.999} & 0.914 & \textbf{0.998} & 0.938 & 0.986 & 0.859 \\
\hdashline
\noalign{\vskip 2pt}
\textbf{R-Log-7B} & 0.995 & \textbf{0.998} & 0.945 & \textbf{0.943} & 0.999 & \textbf{0.993} & 0.996 & \textbf{0.937} & 0.996 & \textbf{0.954} & 0.989 & \textbf{0.922} & 0.991 & \textbf{0.945} & \textbf{0.987} & \textbf{0.956} \\
\bottomrule
\multicolumn{17}{l}{$^{\mathrm{a}}$ \textbf{RI} stands for RandIndex. \textbf{F1} stands for variable-level F1-score.} \\
\end{tabular}
}
\label{tab:logParsing_exp}
\end{table*}

(1) $i=1$ (Log Parsing). This term encourages accurate recognition of both variable and template parts in $Y'$. $\delta_v$ is $1$ if the ground-truth template $Y$ contains variables (\emph{i.e.},“<*>”) and becomes $0$ otherwise. The F1-Score is computed by extracting variables from $Y$ and $Y'$ (treating each variable as a token) and calculating the binary F1-Score with "variable" as the positive label, which evaluates the accuracy of variable extraction in a range of $[0,1]$. For template parts, we use edit distance~\cite{levenshtein1966binary} and normalize it to $[0,1]$ to measure the similarity between the predicted and ground-truth templates.

(2) $i=2$ (Anomaly Detection). The reward value for this sub-task is binary ($\{0,1\}$): it rewards only if the predicted answer $Y'$ is either 'normal' or 'abnormal' and matches the ground truth $Y$. $\mathds{1}$ is the indicator function, reflecting the binary nature of this sub-task.

(3) $i=3,4,5$ (IRS tasks). This term seeks to comprehensively evaluate correctness of semantics in $Y'$ by a combination of BLEU~\cite{papineni2002bleu} and ROUGE~\cite{lin2004rouge} scores. BLEU measures the n-gram (\emph{i.e.}, n consecutive words) precision between $Y$ and $Y'$, emphasizing lexical overlap. ROUGE-1, ROUGE-2, and ROUGE-L measure recall based on unigram (\emph{i.e.}, single word), bigram (\emph{i.e.}, two consecutive words), and longest common subsequence, respectively, assessing content similarity and semantic fluency. These metrics are the most typical ones for comprehensively evaluating semantic correctness in generative text outputs~\cite{wang2020overview,el2021automatic,baradaran2022survey} like the IRS tasks' outputs. The scores are normalized by 400 (the maximum possible sum where each metric is within $[0,100]$).

This joint reward function ensures that the model is penalized for structural errors while being rewarded for semantic accuracy tailored to each sub-task's requirements, facilitating effective learning in the heterogeneous O\&M environment. Note that we didn't impose any reward signal directly on the model-generated reasoning trajectory $R'$, since there might be distinct reasoning paths to reach the reference answers. The initial policy $\theta_c$ is equipped with a human-aligned reasoning paradigm and is expected to self-adapt its strategies according to signals received from the rewards on answers through the RL stage, converging to $\theta_R$, \emph{i.e.}, R-Log.

\section{Experiment}\label{sec:exp}
\subsection{Implementation Details}

Following recent studies~\cite{he2025r1,zhang2025thinkfl}, Qwen2.5-7B~\citep{Yang2024Qwen25TR} is selected as the foundation model $\theta$ due to its steady performance among open-source LLMs. Our main implementation of R-Log, denoted by R-Log-7B, was trained through the two stages according to Eq.~\eqref{eq1} and \eqref{eq2}. For the code-start phase and the implementation of baselines, SFT is performed using the framework of LLaMAFactory~\cite{zheng-etal-2024-llamafactory}. We keep the hyperparameters of SFT the same as LogLM~\cite{liu2025loglm}, training 480 steps with a learning rate of $2\times10^{-5}$ and a batch size of 32. For the RL phase, VeRL~\cite{sheng2024hybridflow} is utilized as the framework. Following He~\emph{et al.}~\cite{he2025r1}, the GRPO training configuration consists of 480 training steps, a learning rate of $4 \times 10^{-7}$, a batch size of 16, and 8 rollouts.

\subsection{Research Questions \& Key Findings}

We empirically studied the following research questions (RQ) and report key findings for each RQ.

\textbf{RQ1:} Can R-Log outperform existing methods, especially LLM-based methods, in diverse log analysis tasks?

\textbf{Key Findings of RQ1:} In Section~\ref{sec:RQ1}, we compare R-Log with various approaches encompassing task-specialized approaches and general-purpose LLMs, on open-source log analysis benchmarks across five sub-tasks. R-Log exhibits strong performance against existing approaches, highlighting its strong application potential in various log analysis scenarios.

\textbf{RQ2:} Does R-Log benefit from its reasoning-based training paradigm with RL?

\textbf{Key Findings of RQ2:} In Section~\ref{sec:RQ2}, we conduct an ablation study on training stages of R-Log. The results indicate that incorporating only the first stage (\emph{i.e.}, SFT with both $R$ and $Y$) improves model performance significantly against plain SFT (only with $Y$), while the RL stage further improves performance over the first stage. This finding substantiates the benefits of our proposed reasoning-based paradigm with RL.

\textbf{RQ3:} Are the human-aligned reasoning strategies important in training R-Log?

\textbf{Key Findings of RQ3:} In Section~\ref{sec:RQ3}, we examine the importance of human reasoning strategies by eliminating the reasoning templates from the dataset construction process (\emph{i.e.}, the reasoning trajectories are generated without any guidance). The result reveals a significant advantage of using the human guidance over free-style generation on task performances, highlighting the importance of human O\&M experience in building reasoning ability for R-Log.

\textbf{RQ4:} Can R-Log generalize to new analysis scenarios?

\textbf{Key Findings of RQ4:} We examine the performance of R-Log and LogLM on a new sub-task unseen from training. The result indicates that R-Log significantly outperforms LogLM in the challenging new scenario, highlighting its strong generalization ability to solve unseen problems via step-by-step reasoning.

\textbf{RQ5:} How to balance between efficacy and efficiency for R-Log?

\textbf{Key Findings of RQ5:} The “think-before-answer” nature of R-Log inevitably sacrifices efficiency due to the long reasoning trajectories. In Section~\ref{sec:RQ5}, we experimented with an interesting alternative: changing nothing except swapping the output order in training phase to be “answer before think”. By using this mode, R-Log first outputs a short answer and can stop when the “<think>” token appears (by setting it as an end token), achieving exact the same speed with non-thinking LLMs. Empirical result suggests only a minor performance drop for the “reversed” version while saving time, indicating a promising trade-off in actual application.

\begin{table*}[tbp]
\caption{Benchmarking with Existing LLMs on Log Interpretation, Root Cause Analysis, and Solution Recommendation.}
\centering
\resizebox{0.84\linewidth}{!}{
\begin{tabular}{@{}l@{\hskip 0.1in}c@{\hskip 0.05in}c@{\hskip 0.05in}c@{\hskip 0.05in}c@{\hskip 0.1in}c@{\hskip 0.05in}c@{\hskip 0.05in}c@{\hskip 0.05in}c@{\hskip 0.1in}c@{\hskip 0.05in}c@{\hskip 0.05in}c@{\hskip 0.05in}c@{}}
\toprule
\multirow{2}{*}{\textbf{Methods}} & \multicolumn{4}{c}{\hspace{-0.4em}\textbf{Log Interpretation}}                              & \multicolumn{4}{c}{\hspace{-1.0em}\textbf{Root Cause Analysis}}                                   & \multicolumn{4}{c}{\hspace{-0.2em}\textbf{Solution Recommendation}}                                   \\ \cmidrule(l{0.3em}r{1.2em}){2-5}  \cmidrule(l{0em}r{1.2em}){6-9} \cmidrule(l{0em}){10-13} 
                                  & \hspace{0.3em}\textbf{BLEU}   & \hspace{0.5em}\textbf{R-1}$^{\mathrm{a}}$ & \textbf{R-2} & \textbf{R-L} & \textbf{BLEU}   & \textbf{R-1} & \textbf{R-2} & \textbf{R-L} & \textbf{BLEU}  & \textbf{R-1} & \textbf{R-2} & \textbf{R-L} \\ \midrule
Qwen2.5-7B-SFT  & 2.041          & 33.852           & 14.140           & 9.289           & 1.605          & 34.210          & 13.586           & 8.873        & 2.032     & 17.474           & 5.588           & 13.084       \\
LLaMA-3.1-405B~\cite{dubey2024llama3}           & 4.466           & 28.563           & 11.106           & 17.071           & 1.416           & 12.984           & 3.694            & 8.162            & 2.018          & 18.768           & 5.287            & 12.045           \\
DeepSeek-V3.1~\cite{deepseekai2024deepseekv3technicalreport}        & 2.162           & 22.274          & 6.353           & 12.754
           & 2.257           & 23.347           & 6.126            & 13.735           & 1.195          & 18.517           & 3.495            & 10.124           \\
Qwen3-235B-A22B~\cite{qwen3technicalreport}                    & 1.858           & 19.974           & 5.949            & 11.148           & 2.329           & 20.355           & 5.936            & 12.318         & 0.916          & 16.159           & 2.964            & 8.879           \\

OWL-7B~\cite{guoowl}                   & 2.566           & 28.211           & 8.289            & 19.123           & 1.947           & 20.893           & 5.671            & 14.718           & 0.953          & 21.620           & 5.006            & 15.574           \\
SuperLog-7B~\cite{ji2025superlog}                   & 4.521           & 32.938           & 12.104            & 20.277           & 3.576           & 27.713           & 8.812            & 17.102           & 3.467          & 27.104           & 7.655            & 17.890           \\
Qwen2.5-7B-Instruct\cite{Yang2024Qwen25TR}  & 7.779          & 39.028           & 14.730           & 28.695           & 7.002          & 35.331          & 15.710           & 26.796        & 3.039     & 24.912           & 8.524           & 18.804       \\
LogLM-7B~\cite{liu2025loglm}                   & 9.826           & 40.462           & 22.545         & 35.810           & 10.281           & 38.741           & 16.803           & 27.524          & 3.970          & 28.083           & 9.332            & 20.183           \\
\hdashline
\noalign{\vskip 2pt}
\textbf{R-Log-7B}              & \textbf{16.671} & \textbf{48.593}  & \textbf{24.754}  & \textbf{43.880}  & \textbf{13.064} & \textbf{41.453}  & \textbf{19.518}  & \textbf{36.492}  & \textbf{6.759} & \textbf{33.587}  & \textbf{12.838}  & \textbf{29.925}  \\ \bottomrule 
\multicolumn{13}{l}{$^{\mathrm{a}}$ \textbf{R-1} stands for ROUGE-1. \textbf{R-2} stands for ROUGE-2. \textbf{R-L} stands for ROUGE-L.}\\
\end{tabular}
}\label{tab:IRS_exp}
\end{table*}

\begin{table}[tbp]
    \caption{Benchmarking on LLM-based Anomaly Detection.}
    \centering
    \resizebox{\linewidth}{!} {%
    \begin{tabular}{l@{\hskip 0.1in}c@{\hskip 0.05in}c@{\hskip 0.05in}c@{\hskip 0.1in}c@{\hskip 0.05in}c@{\hskip 0.05in}c}
    \toprule
    \multirow{2}{*}{\textbf{Methods}} & \multicolumn{3}{c}{\hspace{-1.5em}\textbf{BGL}}       & \multicolumn{3}{c}{\hspace{-0.2em}\textbf{Spirit}} \\ \cmidrule(l{0em}r{1em}){2-4} \cmidrule(l{0em}r{0.2em}){5-7} 
                                   & \hspace{0.6em}\textbf{Pre$^{\mathrm{a}}$} & \textbf{Rec} & \textbf{F1}   & \textbf{Pre} & \textbf{Rec} & \textbf{F1} \\ \midrule
    DeepSeek-V3.1~\cite{deepseekai2024deepseekv3technicalreport} & 0.055 & 0.987 & 0.104 & 0.208 & 0.985 & 0.344 \\
    Qwen3-235B-A22B~\cite{qwen3technicalreport}  & 0.050 & \textbf{1.000} & 0.095 & 0.195 & \textbf{1.000} & 0.327 \\
    OWL-7B~\cite{guoowl}      & 0.081 & 0.197 & 0.115 & 0.230 & 0.943 & 0.221 \\
    SuperLog-7B~\cite{ji2025superlog}      & 0.385 & 0.197 & 0.261 & \textbf{0.778} & 0.051 & 0.097 \\
    Qwen2.5-7B-SFT  & 0.429 & 0.723 & 0.539 & 0.314 & 0.699 & 0.433 \\
    LogLM-7B~\cite{liu2025loglm}            & 0.400 & 0.605 & 0.482 & 0.429 & 0.728 & 0.540 \\
    PreLog~\cite{le2024prelog}    & 0.426 & 0.605 & 0.500 & 0.531 & 0.191 & 0.281 \\
    LogPrompt~\cite{liu2024logprompt}      & 0.447 & 0.829 & 0.581 & 0.402 & 0.794 & 0.533 \\
    LogGPT~\cite{qi2023loggpt} & 0.056 & 0.974 & 0.106 & 0.131 & 0.985 & 0.232 \\
    \hdashline
    \noalign{\vskip 2pt}
    \textbf{R-Log-7B}        & \textbf{0.493} & 0.908 & \textbf{0.639} & 0.446 & 0.934 & \textbf{0.603} \\ 
    \bottomrule 
    \multicolumn{7}{l}{$^{\mathrm{a}}$ \textbf{Pre}, \textbf{Rec}, \textbf{F1} stands for Precision, Recall and F1-score.} \\
    \end{tabular}
    }
    \label{tab:anomaly_exp}
\end{table}

\subsection{RQ1: Benchmarking on Log Analysis Tasks}\label{sec:RQ1}

\subsubsection{Evaluation Datasets}

We evaluate the performance of R-Log using the open-source benchmark from LogLM~\cite{liu2025loglm}, encompassing the five log analysis sub-tasks and human-inspected reference answers. All the logs in the test suites are from real-world scenarios, encompassing 10 distinct domains. For log parsing and anomaly detection, the templates~\cite{he2020loghub} and anomaly labels~\cite{oliner2007supercomputers} are manually annotated by domain experts. In addition, we applied the improved annotations for anomaly detection from Xu~\emph{et al.}~\cite{xu2025rationanomalyloganomalydetection}, who identified and corrected around 10\% of the annotation errors in the original datasets by \cite{oliner2007supercomputers}. For the IRS tasks, the logs are from user posts in technical forums and the reference answers are the highest voted answers~\cite{wang2024logexpert} with human examinations by \cite{liu2025loglm}. The ratio between training samples and testing samples is 1:9 for log parsing and anomaly detection and 8:2 for IRS tasks, with strict separation between the sets to avoid data leakage~\cite{liu2025loglm}. All baselines (if trainable) were trained using the same amount of task-specific training data to ensure fair comparison.

\subsubsection{Baselines}

We compare R-Log with the following three groups of baselines for comprehensiveness: \textbf{(1) Task-specific methods}, including algorithms and models targeted for a single sub-task such as LogGPT~\cite{qi2023loggpt}, Drain~\cite{he2017drain}, LILAC~\cite{jiang2024lilac} and LogStamp~\cite{tao2022logstamp}; \textbf{(2) General-purpose LLMs}, including various powerful LLMs with strong language processing ability such as DeepSeek-V3.1~\cite{deepseekai2024deepseekv3technicalreport}, LLaMA-3.1-405B~\cite{dubey2024llama3} and Qwen series~\cite{qwen3technicalreport,Yang2024Qwen25TR}; \textbf{(3) Domain LLMs}, including LLMs that incorporated domain knowledge for multiple sub-tasks within log analysis. This category includes LogPrompt~\cite{liu2024logprompt} which drives LLMs with specially-designed prompts for log analysis, PreLog~\cite{le2024prelog}, OWL-7B~\cite{guoowl} and SuperLog-7B~\cite{ji2025superlog} which are domain-adapted LLMs pre-trained with data related to IT operation and log analysis, and LogLM-7B~\cite{liu2025loglm} which utilizes instruction tuning to build multi-task log analysis ability for LLMs. We reproduced these models to ensure a fair comparison under the same experimental settings (\emph{e.g.}, foundation model). In addition, we also fine-tuned the foundation model $\theta$ with the same in-domain data as R-Log for each sub-task, denoted by Qwen-2.5-SFT.

\subsubsection{Log Parsing}\label{sec:RQ1_parsing}

Table~\ref{tab:logParsing_exp} displays the benchmarking result for log parsing. Following existing studies~\emph{et al.}~\cite{meng2020logparse,tao2023biglog,tao2022logstamp,liu2025loglm}, RandIndex~\cite{rand1971objective} and variable-level F1-score~\cite{liu2024interpretable} are utilized as the metric for measuring the accurate recognition of template and variable, respectively. RandIndex assesses the accuracy of log clustering (\emph{i.e.}, whether two logs with the same template are accurately clustered together), regardless of the correctness of variables in the extracted templates. In contrast, F1-score focuses on successful identification of variables in logs, thereby serving as a more challenging metric. The result indicates that R-Log achieves a strong performance (best averagely) on both template extraction and variable recognition, compared with existing methods which may struggle on this online scenario where majority of the logs are unseen from training (\emph{i.e.}, a training-test ratio of 1:9).

\subsubsection{Anomaly Detection}\label{sec:RQ1_anomaly}

We report F1-score of anomalies as the metric for a comprehensive evaluation on the sub-task of anomaly detection. The F1-score is computed as the harmonic average of Precision and Recall, where Precision measures the percentage of correctly classified ones in all of model's predicted anomalies, and Recall measures the recall rate of golden abnormal logs. An imbalance on Precision and Recall reflects failure of the system in anomaly detection and will lead to low F1-score. For instance, in Table~\ref{tab:anomaly_exp}, the high Recall and low Precision of Qwen3-235B-A22B means it classified most logs as abnormal, while the low Recall and high Precision of SuperLog in the dataset of Spirit means most logs are classified as normal. In contrast, R-Log achieves the highest F1-score, indicating its practical ability in detecting anomalies.

\subsubsection{IRS Tasks}\label{sec:RQ1_IRS}

As shown in Table~\ref{tab:IRS_exp}, R-Log continuously outperforms existing LLMs in both BLEU and Rouge scores. The advantages in BLEU indicates a more precise and readable answer with less hallucinated contents, while the high Rouge scores suggest a better recall of the key points in reference answers, such as failure causes in the sub-task of root cause analysis and mitigating steps in the sub-task of solution recommendation.

\subsection{RQ2: Ablation Study on Training Stages}\label{sec:RQ2}

\begin{figure}[tbp]
 \centering  
 \subfigbottomskip=-2pt 
 \subfigcapskip=-2pt 
 \subfigure[Stage-wise Ablation]{
  \includegraphics[width=0.967\linewidth]{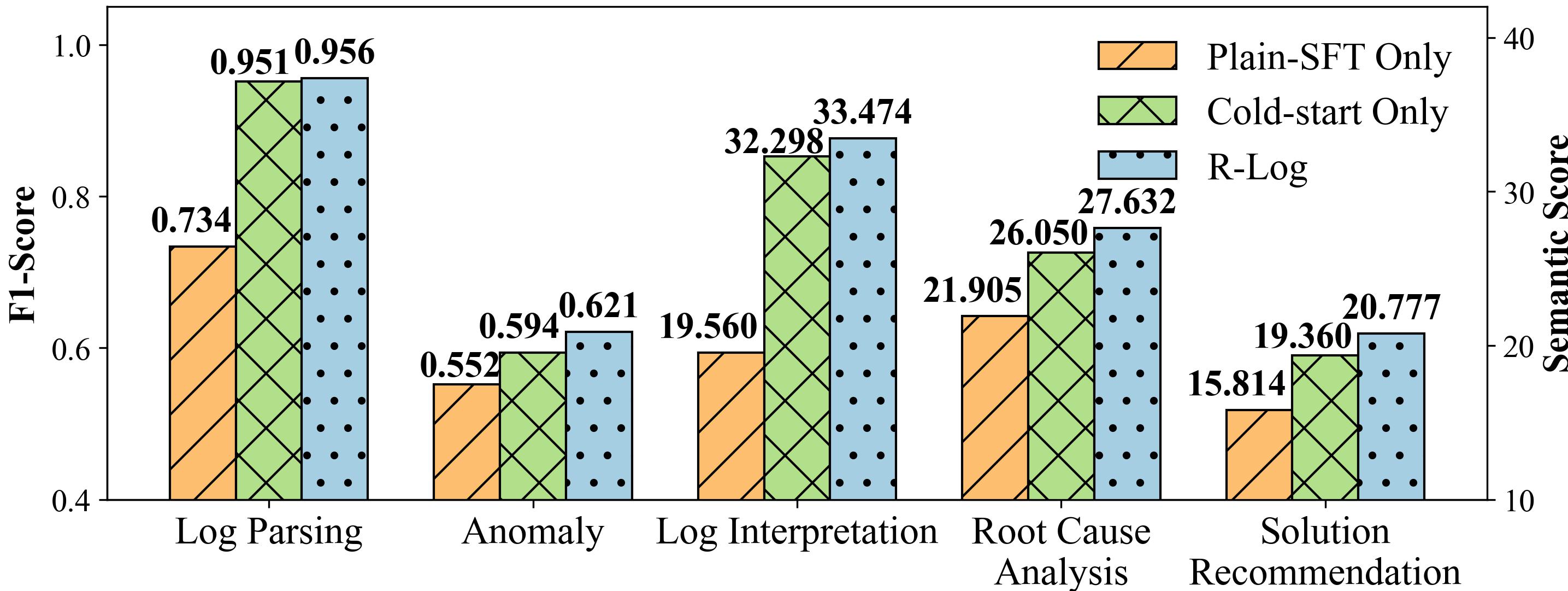}}
   \\
 \subfigure[Ablation on Human Guidance]{
  \includegraphics[width=0.96\linewidth]{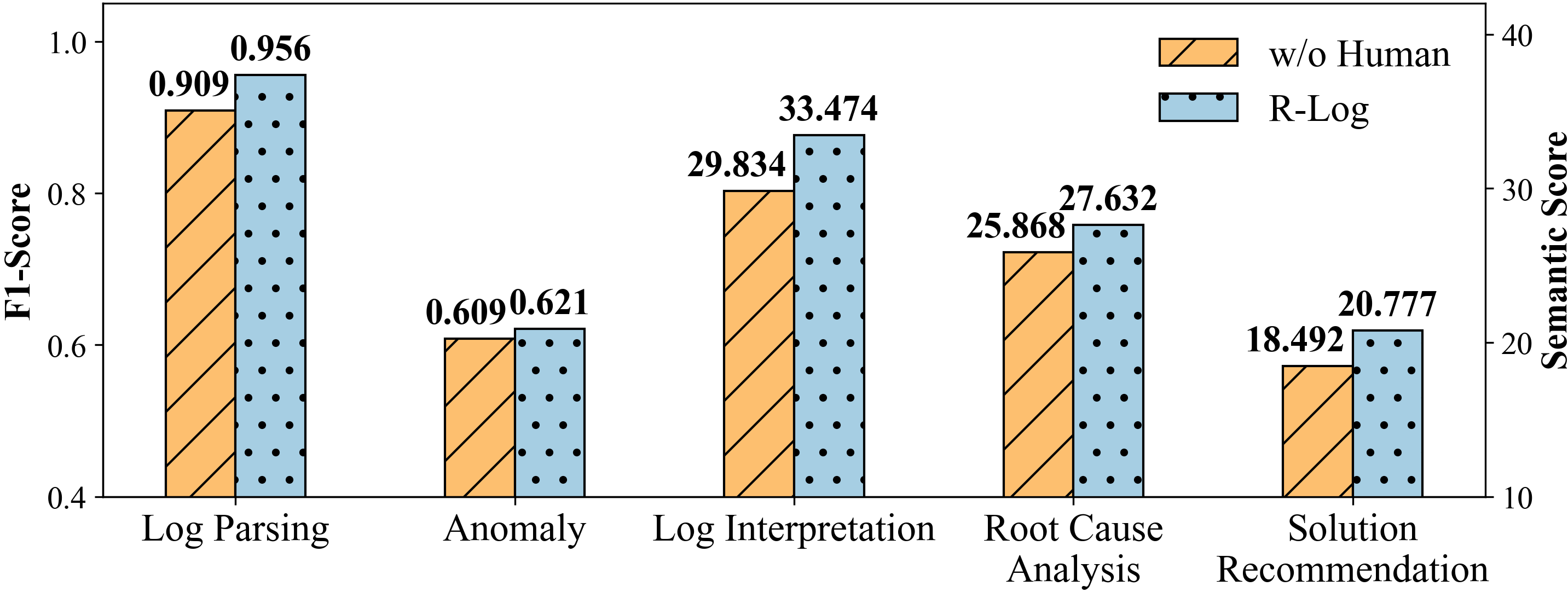}}
 \caption{Ablation study on (a) training stages and (b) human reasoning templates. Scores are averaged across domains.}
\label{fig_ablation}
\end{figure}

For this ablation study, three LLMs are compared under the same experimental settings, except the following differences: \textbf{(1) R-Log:} The exact R-Log-7B model with full cold-start and RL stages; \textbf{(2) Cold-start only:} This model didn't go through the RL stage and was only fine-tuned on Log Reasoning Dataset with (X,R,Y) samples; \textbf{(3) Plain-SFT only:} This model is trained with the same Log Reasoning Dataset, but only on (X,Y) samples without the reasoning trajectories. The scores reported in Fig.~\ref{fig_ablation} are averaged across multiple evaluation datasets for the five sub-tasks (\emph{e.g.}, averaging BGL and Spirit for anomaly detection). For log parsing and anomaly detection, we report the F1-score; for IRS tasks, we report the average the four semantic scores. \textbf{All models are aligned in its number of trained tokens} (by upsampling the training set of (2) and (3)) to offset benefits by just seeing more data in training.

As shown in Fig.~\ref{fig_ablation}(a), the advantage of “Cold-start only” over “Plain-SFT only” suggests the benefits of introducing reasoning-based paradigms into LLMs, where the step-by-step reasoning process facilitates the learning of problem-solving abilities in complex log analysis tasks compared with fitting only on the log-label pairs. However, as discussed in Section~\ref{sec:intro}, SFT flattens the supervision signals to every word (\emph{i.e.}, $R$ and $Y$). Since $R$ is significantly longer than $Y$, the model may fail to optimize its answer quality. After the RL stages, the model performance continuously advances on the five sub-tasks, suggesting the effectiveness of the reward signals on answers where the model further optimizes its reasoning paradigms by handling log events in a heterogeneous O\&M environment. The improvement by RL in log parsing is less significant than other four sub-tasks, probably due to the relative easier nature of the task.

\subsection{RQ3: Ablation Study on Human Strategies}\label{sec:RQ3}

We compare the following two groups: \textbf{(1) R-Log:} The exact RL-enhanced R-Log-7B model cold-started with human-aligned reasoning paradigm; \textbf{(2) W/o human:} The same setting with R-Log, with only difference in cold-start stage. Human-aligned reasoning guidance (\emph{i.e.}, the templates) are removed from the prompt for instantiation of reasoning trajectories, resulting in a free-style generation of reasoning trajectories reflecting the LLM's paradigm.

As shown in Fig.~\ref{fig_ablation}(b), without the incorporation of human-aligned reasoning templates, the performance continuously degrades in all sub-tasks. These reasoning templates comprehensively reflect diverse thinking strategies human engineers adopt in O\&M practices, leading to high-quality reasoning trajectories with less diverted or hallucinated content than those purely generated by LLMs. These trajectories then provide a initial reasoning paradigm for R-Log, facilitating its further optimization through RL.  

\subsection{RQ4: Generalization on Unseen Task}\label{sec:RQ4}

\begin{table}[tbp]
    \caption{F1-score of Variable Categories in an Unseen Task.}
    \centering
    \resizebox{0.9\linewidth}{!}{
    \begin{tabular}{l@{\hskip 0.1in}c@{\hskip 0.05in}c@{\hskip 0.05in}c@{\hskip 0.05in}c@{\hskip 0.05in}c@{\hskip 0.05in}c}
    \toprule
    \textbf{Methods} & \textbf{OID$^{\mathrm{a}}$} & \textbf{LOI} & \textbf{OBN} & \textbf{TDA} & \textbf{CRS} & \textbf{OBA} \\ 
    \midrule
    LogLM-7B~\cite{liu2025loglm} & 0.226 & 0.269 & 0.037 & 0.166 & 0.101 & 0.031 \\
\hdashline
\noalign{\vskip 3pt}
 \multicolumn{7}{l}{\textbf{R-Log-7B}} \\
    \textit{w/o RL} & 0.292 & 0.332 & 0.037 & 0.206 & 0.041 & 0.112 \\
    \textbf{with RL} & \textbf{0.567} & \textbf{0.665} & \textbf{0.188} & \textbf{0.714} & \textbf{0.244} & \textbf{0.344} \\ 
    \bottomrule 
   \multicolumn{7}{l}{$^{\mathrm{a}}$ \textbf{OID}, \textbf{LOI}, \textbf{OBN}, \textbf{TDA}, \textbf{CRS}, \textbf{OBA} are variable classes. } \\
    \end{tabular}
    }
    \label{tab:unseen task}
\end{table}

\begin{figure}[tbp]
    \centering
  \includegraphics[width=0.9\linewidth]{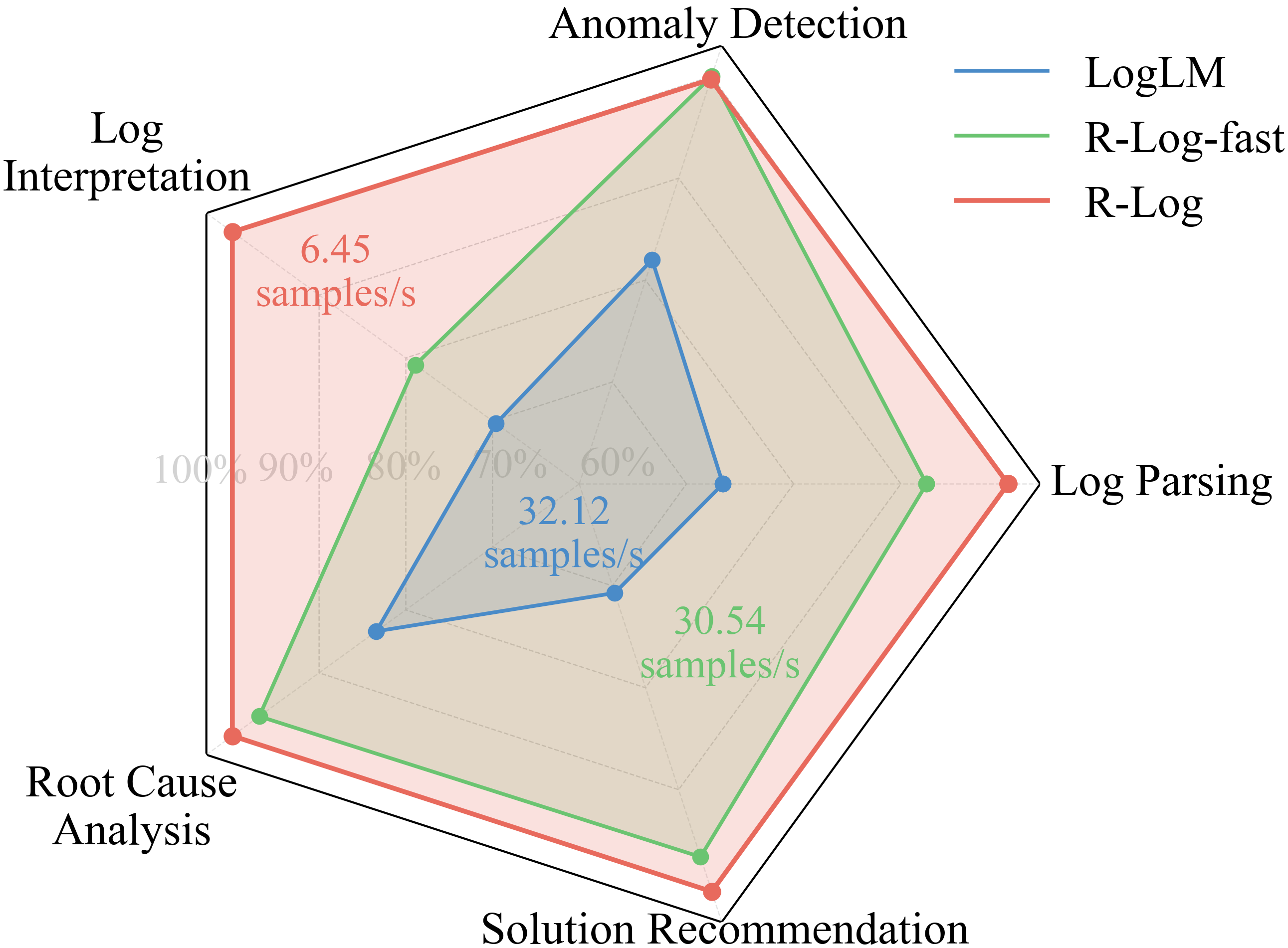}
  \caption{“Think-before-answer” (R-Log) \emph{v.s.} “Answer-before-think” (R-Log-fast), for a trade-off between efficacy and efficiency. The radar displays the relative percentage of baselines' average performances in comparison to R-Log's.}
  \label{fig_reverse}
\end{figure}

\begin{figure*}[tbp]
    \centering
  \includegraphics[width=\linewidth]{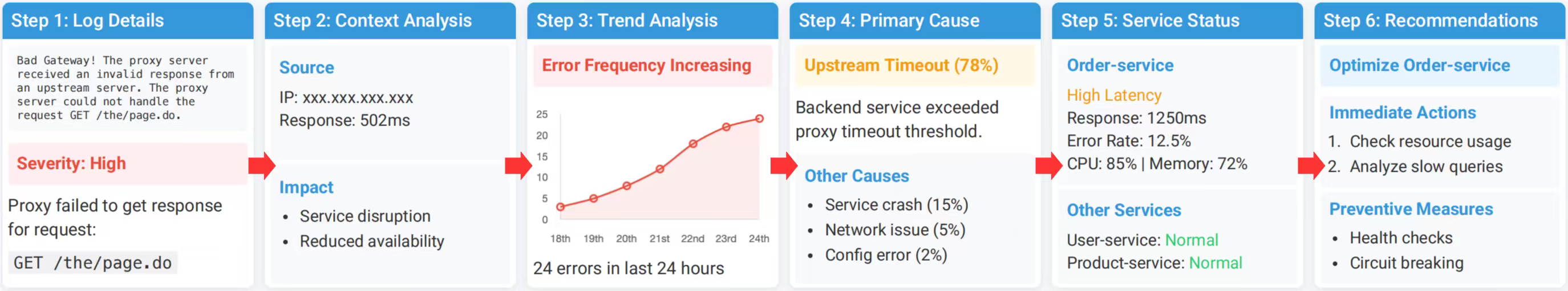}
  \caption{R-Log is deployed in an auto-troubleshooting application emphasizing interpretability in Huawei. The generated cards visualize reasoning trajectories of R-Log handling logged errors. The user data presented is synthesized for demo purpose.}
  \label{fig_in_practice}
\end{figure*}

We excluded \textit{Log Variable Classification}, a challenging sub-task requiring extensive log-related knowledge as described in Section~\ref{sec:log task intro}, from the training phase of R-Log to examine its generalization ability. The test set comprises 14000 logs from seven domains. Li~\emph{et al.}~\cite{li2023did} manually classified each variable within these logs into the category of Object ID (OID), Location Indicator (LOI), Object Name (OBN), Time/Duration of an Action (TDA), Computing Resources (CRS) and Object Amount (OBA), leading to a total of 29711 annotations of variable category. Our task instruction contains short descriptions of these categories and a random example as format guidance. We report F1-score of these categories as a metric for accurate classification of variables.

As shown in Table~\ref{tab:unseen task}, despite not trained with any labels on variable categories, R-Log still demonstrates a strong performance compared with LogLM, especially on the categories of OID (0.567 \emph{v.s.} 0.226), LOI (0.665 \emph{v.s.} 0.269) and TDA (0.714 \emph{v.s.} 0.166). Without training on the categories, the model must deduce variable types purely from the log context. By decomposing complex problems into intermediate steps through a step-by-step, human-like reasoning paradigm, R-Log generalizes its log analysis knowledge more effectively, leading to its advantage on this new task. An interesting finding is that RL stage incentivizes reasoning abilities significantly more in this challenging scenario than in seen tasks (\emph{i.e.}, Fig.~\ref{fig_ablation}), suggesting the vital role RL plays in avoiding mere memorization.

\subsection{RQ5: Fastening via Answer-before-thinking}\label{sec:RQ5}

A recent study~\cite{kudo2024think} on reasoning-based LLMs revealed an interesting phenomenon: the model may already “know” the answers at the beginning of its reasoning process. This inspired us to design a fastening version of R-Log (denoted as R-Log-fast), by switching the position of $R$ and $Y$ in the training dataset to induce an answer-first model. R-Log-fast first outputs answer $Y$, then outputs $R$, which can be intercepted by stopping at the “<think>” token, thereby avoiding the lengthy $R$ and achieving the same speed as non-thinking LLMs. 

As shown in Fig.~\ref{fig_reverse}, the performance reduction of R-Log-fast is tolerable compared with its efficiency enhancement (nearly 5x faster). Despite using no reasoning at inference time, R-Log-fast, trained on full reasoning data with RL, still significantly outperforms LogLM, makes it a promising candidate for online deployments where latency is critical. We also noted the relatively significant performance reduction of R-Log-fast on log interpretation, which is consistent with Fig.~\ref{fig1}, where small but critical details like "1ms" in this sub-task are easily missed without fully step-by-step reasoning.

\section{Discussion}
\subsection{R-Log in practice}\label{sec:in practice}

Given R-Log's strong analysis ability and unique white-box paradigm which outputs reasoning trajectories to reach the conclusion, it has been deployed in the next-gen auto-troubleshooting application within a software and network O\&M platform of Huawei. The application generates a series of vivid card-like UI (\emph{e.g.}, containing a line chart showing invocation trends) as its core feature to visualize the log-based troubleshooting process. Each card represents an intermediate step, making the analysis conclusions more interpretable to users and improving the overall user experience. R-Log serves as the backend model, where its step-by-step reasoning trajectories are further visualized combining code-based UI generation techniques~\cite{UICopilot} into a series of HTML-based cards. To support retrieving real-time logs and data from local devices via private API, private data involving function calling~\cite{schick2024toolformer} were also added to the training dataset of the deployed version of R-Log. Once the error log is triggered, R-Log first invokes API to collect related device or service data, and conduct its reasoning process until reaching the conclusion. The entire reasoning trajectory is used as input for a UI model, which generates multiple cards using HTML language, each displaying a different part of the process.

Fig.~\ref{fig_in_practice} shows an illustrative example of the generated troubleshooting cards, visualizing R-Log's analysis process of a bad gateway error with six cards. The model begins by parsing the raw log, extracting key variables (GET /the/page.do) (card 1), and contextualizing the event by identifying a sharp surge in frequency—24 identical errors within 24 hours (card 2 and 3). It then deduces Upstream Service Timeout (78\% confidence) as the root cause (card 4), validated by correlating the failure with severe latency (1250ms) and high error rates (12.5\%) specifically from the order-service dependency (card 5). Finally (card 6), R-Log concludes by synthesizing these insights into a series of remediation strategy to optimize the order-service. By visualizing the troubleshooting process into a series of vivid cards with diagnostic narrative and operational guidance, this application based on R-Log accelerates mitigation workflows with highly interpretable reasoning trajectories. Since its deployment, R-Log has processed over 100k queries, averagely 500+ queries per day.

\subsection{Threats to Validity}

Our study has several limitations: 
\textbf{(1) Extra Latency by Reasoning:} Reasoning-based LLMs introduce extra latency that may burden online log analysis~\cite{ma2025practitioners}. To alleviate this, we propose R-Log-fast with an "Answer-first" strategy, fastening the inference while preserving performance, thereby achieving a practical balance between efficiency and accuracy.
\textbf{(2) Limited Size of Foundation Model:} Due to tight budget of resources, we didn't verify R-Log using larger foundation models, reducing the confidence of results. However, 7B is the most popular model size~\cite{liu2025fin,he2025r1,chu2025domaino1s} for reasoning-based LLMs which balances between accuracy and speed in deployment.
\textbf{(3) Fairness of Experimental Comparisons:} In Section~\ref{sec:RQ1}, discrepancies between R-Log's training data and baselines raised fairness concerns. However, fully unifying data is impractical due to proprietary, pre-trained baselines. To ensure fairness, in Section~\ref{sec:RQ2}, we applied token alignment to offset extra-data effects.

\section{Conclusion}
In this paper, we present R-Log, a model that redefines LLM-based log analysis through a reasoning-first learning paradigm. By using RL with reasoning trajectories instead of SFT, R-Log learns the underlying rules of analysis. This method enables SOTA performance on known tasks and, more importantly, exceptional generalization to unseen problems by decomposing them into logical steps. This capability, combined with the verifiable trust offered by its transparent reasoning, was instrumental in its successful deployment for managing complex software systems. Future work include testing on more datasets, tasks and larger models, and refining rewards.

\bibliographystyle{ACM-Reference-Format}
\bibliography{mybib}


\begin{thebibliography}{72}


\ifx \showCODEN    \undefined \def \showCODEN     #1{\unskip}     \fi
\ifx \showISBNx    \undefined \def \showISBNx     #1{\unskip}     \fi
\ifx \showISBNxiii \undefined \def \showISBNxiii  #1{\unskip}     \fi
\ifx \showISSN     \undefined \def \showISSN      #1{\unskip}     \fi
\ifx \showLCCN     \undefined \def \showLCCN      #1{\unskip}     \fi
\ifx \shownote     \undefined \def \shownote      #1{#1}          \fi
\ifx \showarticletitle \undefined \def \showarticletitle #1{#1}   \fi
\ifx \showURL      \undefined \def \showURL       {\relax}        \fi
\providecommand\bibfield[2]{#2}
\providecommand\bibinfo[2]{#2}
\providecommand\natexlab[1]{#1}
\providecommand\showeprint[2][]{arXiv:#2}

\bibitem[Ahmed et~al\mbox{.}(2023)]%
        {ahmed2023recommending}
\bibfield{author}{\bibinfo{person}{Toufique Ahmed}, \bibinfo{person}{Supriyo Ghosh}, \bibinfo{person}{Chetan Bansal}, \bibinfo{person}{Thomas Zimmermann}, \bibinfo{person}{Xuchao Zhang}, {and} \bibinfo{person}{Saravan Rajmohan}.} \bibinfo{year}{2023}\natexlab{}.
\newblock \showarticletitle{Recommending Root-Cause and Mitigation Steps for Cloud Incidents using Large Language Models}. In \bibinfo{booktitle}{\emph{ICSE 2023}}.
\newblock


\bibitem[Baradaran et~al\mbox{.}(2022)]%
        {baradaran2022survey}
\bibfield{author}{\bibinfo{person}{Razieh Baradaran}, \bibinfo{person}{Razieh Ghiasi}, {and} \bibinfo{person}{Hossein Amirkhani}.} \bibinfo{year}{2022}\natexlab{}.
\newblock \showarticletitle{A survey on machine reading comprehension systems}.
\newblock \bibinfo{journal}{\emph{Natural Language Engineering}} \bibinfo{volume}{28}, \bibinfo{number}{6} (\bibinfo{year}{2022}), \bibinfo{pages}{683--732}.
\newblock


\bibitem[Botvinick et~al\mbox{.}(2020)]%
        {botvinick2020deep}
\bibfield{author}{\bibinfo{person}{Matthew Botvinick}, \bibinfo{person}{Jane~X Wang}, \bibinfo{person}{Will Dabney}, \bibinfo{person}{Kevin~J Miller}, {and} \bibinfo{person}{Zeb Kurth-Nelson}.} \bibinfo{year}{2020}\natexlab{}.
\newblock \showarticletitle{Deep reinforcement learning and its neuroscientific implications}.
\newblock \bibinfo{journal}{\emph{Neuron}} \bibinfo{volume}{107}, \bibinfo{number}{4} (\bibinfo{year}{2020}), \bibinfo{pages}{603--616}.
\newblock


\bibitem[Chen et~al\mbox{.}(2024)]%
        {chen2024automatic}
\bibfield{author}{\bibinfo{person}{Yinfang Chen}, \bibinfo{person}{Huaibing Xie}, \bibinfo{person}{Minghua Ma}, \bibinfo{person}{Yu Kang}, \bibinfo{person}{Xin Gao}, \bibinfo{person}{Liu Shi}, \bibinfo{person}{Yunjie Cao}, \bibinfo{person}{Xuedong Gao}, \bibinfo{person}{Hao Fan}, \bibinfo{person}{Ming Wen}, {et~al\mbox{.}}} \bibinfo{year}{2024}\natexlab{}.
\newblock \showarticletitle{Automatic root cause analysis via large language models for cloud incidents}. In \bibinfo{booktitle}{\emph{Proceedings of the Nineteenth European Conference on Computer Systems}}. \bibinfo{pages}{674--688}.
\newblock


\bibitem[Chu et~al\mbox{.}(2025b)]%
        {chusft}
\bibfield{author}{\bibinfo{person}{Tianzhe Chu}, \bibinfo{person}{Yuexiang Zhai}, \bibinfo{person}{Jihan Yang}, \bibinfo{person}{Shengbang Tong}, \bibinfo{person}{Saining Xie}, \bibinfo{person}{Dale Schuurmans}, \bibinfo{person}{Quoc~V Le}, \bibinfo{person}{Sergey Levine}, {and} \bibinfo{person}{Yi Ma}.} \bibinfo{year}{2025}\natexlab{b}.
\newblock \showarticletitle{SFT Memorizes, RL Generalizes: A Comparative Study of Foundation Model Post-training}. In \bibinfo{booktitle}{\emph{Forty-second International Conference on Machine Learning}}.
\newblock


\bibitem[Chu et~al\mbox{.}(2025a)]%
        {chu2025domaino1s}
\bibfield{author}{\bibinfo{person}{Xu Chu}, \bibinfo{person}{Zhijie Tan}, \bibinfo{person}{Hanlin Xue}, \bibinfo{person}{Guanyu Wang}, \bibinfo{person}{Tong Mo}, {and} \bibinfo{person}{Weiping Li}.} \bibinfo{year}{2025}\natexlab{a}.
\newblock \showarticletitle{Domaino1s: Guiding llm reasoning for explainable answers in high-stakes domains}.
\newblock \bibinfo{journal}{\emph{arXiv preprint arXiv:2501.14431}} (\bibinfo{year}{2025}).
\newblock


\bibitem[Curtis(1984)]%
        {curtis1984fifteen}
\bibfield{author}{\bibinfo{person}{Bill Curtis}.} \bibinfo{year}{1984}\natexlab{}.
\newblock \showarticletitle{Fifteen years of psychology in software engineering: Individual differences and cognitive science}. In \bibinfo{booktitle}{\emph{Proceedings of the 7th international conference on Software engineering}}. \bibinfo{pages}{97--106}.
\newblock


\bibitem[DeepSeek-AI(2024)]%
        {deepseekai2024deepseekv3technicalreport}
\bibfield{author}{\bibinfo{person}{DeepSeek-AI}.} \bibinfo{year}{2024}\natexlab{}.
\newblock \bibinfo{title}{DeepSeek-V3 Technical Report}.
\newblock
\showeprint[arxiv]{2412.19437}~[cs.CL]
\urldef\tempurl%
\url{https://arxiv.org/abs/2412.19437}
\showURL{%
\tempurl}


\bibitem[DeepSeek-AI(2025)]%
        {deepseek2025r1}
\bibfield{author}{\bibinfo{person}{DeepSeek-AI}.} \bibinfo{year}{2025}\natexlab{}.
\newblock \showarticletitle{DeepSeek-R1: Incentivizing Reasoning Capability in LLMs via Reinforcement Learning}. In \bibinfo{booktitle}{\emph{arXiv preprint arXiv:2501.12948}}.
\newblock


\bibitem[Du and Li(2016)]%
        {du2016spell}
\bibfield{author}{\bibinfo{person}{Min Du} {and} \bibinfo{person}{Feifei Li}.} \bibinfo{year}{2016}\natexlab{}.
\newblock \showarticletitle{Spell: Streaming parsing of system event logs}. In \bibinfo{booktitle}{\emph{2016 IEEE 16th International Conference on Data Mining (ICDM)}}. IEEE, \bibinfo{pages}{859--864}.
\newblock


\bibitem[Du et~al\mbox{.}(2017)]%
        {du2017deeplog}
\bibfield{author}{\bibinfo{person}{Min Du}, \bibinfo{person}{Feifei Li}, \bibinfo{person}{Guineng Zheng}, {and} \bibinfo{person}{Vivek Srikumar}.} \bibinfo{year}{2017}\natexlab{}.
\newblock \showarticletitle{Deeplog: Anomaly detection and diagnosis from system logs through deep learning}. In \bibinfo{booktitle}{\emph{Proceedings of the 2017 ACM SIGSAC conference on computer and communications security}}. \bibinfo{pages}{1285--1298}.
\newblock


\bibitem[Dubey et~al\mbox{.}(2024)]%
        {dubey2024llama3}
\bibfield{author}{\bibinfo{person}{Abhimanyu Dubey}, \bibinfo{person}{Abhinav Jauhri}, \bibinfo{person}{Abhinav Pandey}, \bibinfo{person}{Abhishek Kadian}, \bibinfo{person}{Ahmad Al-Dahle}, \bibinfo{person}{Aiesha Letman}, \bibinfo{person}{Akhil Mathur}, \bibinfo{person}{Alan Schelten}, \bibinfo{person}{Amy Yang}, \bibinfo{person}{Angela Fan}, {et~al\mbox{.}}} \bibinfo{year}{2024}\natexlab{}.
\newblock \showarticletitle{The Llama 3 Herd of Models}.
\newblock \bibinfo{journal}{\emph{arXiv preprint arXiv:2407.21783}} (\bibinfo{year}{2024}).
\newblock


\bibitem[El-Kassas et~al\mbox{.}(2021)]%
        {el2021automatic}
\bibfield{author}{\bibinfo{person}{Wafaa~S El-Kassas}, \bibinfo{person}{Cherif~R Salama}, \bibinfo{person}{Ahmed~A Rafea}, {and} \bibinfo{person}{Hoda~K Mohamed}.} \bibinfo{year}{2021}\natexlab{}.
\newblock \showarticletitle{Automatic text summarization: A comprehensive survey}.
\newblock \bibinfo{journal}{\emph{Expert systems with applications}}  \bibinfo{volume}{165} (\bibinfo{year}{2021}), \bibinfo{pages}{113679}.
\newblock


\bibitem[Fu et~al\mbox{.}(2009)]%
        {fu2009execution}
\bibfield{author}{\bibinfo{person}{Qiang Fu}, \bibinfo{person}{Jian-Guang Lou}, \bibinfo{person}{Yi Wang}, {and} \bibinfo{person}{Jiang Li}.} \bibinfo{year}{2009}\natexlab{}.
\newblock \showarticletitle{Execution anomaly detection in distributed systems through unstructured log analysis}. In \bibinfo{booktitle}{\emph{2009 ninth IEEE international conference on data mining}}. \bibinfo{pages}{149--158}.
\newblock


\bibitem[Gui et~al\mbox{.}(2025)]%
        {UICopilot}
\bibfield{author}{\bibinfo{person}{Yi Gui}, \bibinfo{person}{Yao Wan}, \bibinfo{person}{Zhen Li}, \bibinfo{person}{Zhongyi Zhang}, \bibinfo{person}{Dongping Chen}, \bibinfo{person}{Hongyu Zhang}, \bibinfo{person}{Yi Su}, \bibinfo{person}{Bohua Chen}, \bibinfo{person}{Xing Zhou}, \bibinfo{person}{Wenbin Jiang}, {and} \bibinfo{person}{Xiangliang Zhang}.} \bibinfo{year}{2025}\natexlab{}.
\newblock \showarticletitle{UICopilot: Automating UI Synthesis via Hierarchical Code Generation from Webpage Designs}. In \bibinfo{booktitle}{\emph{Proceedings of the ACM on Web Conference 2025}} (Sydney NSW, Australia) \emph{(\bibinfo{series}{WWW '25})}. \bibinfo{publisher}{Association for Computing Machinery}, \bibinfo{address}{New York, NY, USA}, \bibinfo{pages}{1846–1855}.
\newblock
\showISBNx{9798400712746}
\href{https://doi.org/10.1145/3696410.3714891}{doi:\nolinkurl{10.1145/3696410.3714891}}


\bibitem[Guideline(1992)]%
        {guideline1992root}
\bibfield{author}{\bibinfo{person}{DOE Guideline}.} \bibinfo{year}{1992}\natexlab{}.
\newblock \showarticletitle{Root cause analysis guidance document}.
\newblock \bibinfo{journal}{\emph{US Department of Energy: Washington}} (\bibinfo{year}{1992}).
\newblock


\bibitem[Guo et~al\mbox{.}(2024)]%
        {guoowl}
\bibfield{author}{\bibinfo{person}{Hongcheng Guo}, \bibinfo{person}{Jian Yang}, \bibinfo{person}{Jiaheng Liu}, \bibinfo{person}{Liqun Yang}, \bibinfo{person}{Linzheng Chai}, \bibinfo{person}{Jiaqi Bai}, \bibinfo{person}{Junran Peng}, \bibinfo{person}{Xiaorong Hu}, \bibinfo{person}{Chao Chen}, \bibinfo{person}{Dongfeng Zhang}, {et~al\mbox{.}}} \bibinfo{year}{2024}\natexlab{}.
\newblock \showarticletitle{OWL: A Large Language Model for IT Operations}. In \bibinfo{booktitle}{\emph{The Twelfth International Conference on Learning Representations}}.
\newblock


\bibitem[He et~al\mbox{.}(2025)]%
        {he2025r1}
\bibfield{author}{\bibinfo{person}{Minggui He}, \bibinfo{person}{Yilun Liu}, \bibinfo{person}{Shimin Tao}, \bibinfo{person}{Yuanchang Luo}, \bibinfo{person}{Hongyong Zeng}, \bibinfo{person}{Chang Su}, \bibinfo{person}{Li Zhang}, \bibinfo{person}{Hongxia Ma}, \bibinfo{person}{Daimeng Wei}, \bibinfo{person}{Weibin Meng}, {et~al\mbox{.}}} \bibinfo{year}{2025}\natexlab{}.
\newblock \showarticletitle{R1-t1: Fully incentivizing translation capability in llms via reasoning learning}.
\newblock \bibinfo{journal}{\emph{arXiv preprint arXiv:2502.19735}} (\bibinfo{year}{2025}).
\newblock


\bibitem[He et~al\mbox{.}(2017)]%
        {he2017drain}
\bibfield{author}{\bibinfo{person}{Pinjia He}, \bibinfo{person}{Jieming Zhu}, \bibinfo{person}{Zibin Zheng}, {and} \bibinfo{person}{Michael~R Lyu}.} \bibinfo{year}{2017}\natexlab{}.
\newblock \showarticletitle{Drain: An online log parsing approach with fixed depth tree}. In \bibinfo{booktitle}{\emph{2017 IEEE international conference on web services (ICWS)}}. IEEE, \bibinfo{pages}{33--40}.
\newblock


\bibitem[He et~al\mbox{.}(2020)]%
        {he2020loghub}
\bibfield{author}{\bibinfo{person}{Shilin He}, \bibinfo{person}{Jieming Zhu}, \bibinfo{person}{Pinjia He}, {and} \bibinfo{person}{Michael~R Lyu}.} \bibinfo{year}{2020}\natexlab{}.
\newblock \showarticletitle{Loghub: a large collection of system log datasets towards automated log analytics}.
\newblock \bibinfo{journal}{\emph{arXiv preprint arXiv:2008.06448}} (\bibinfo{year}{2020}).
\newblock


\bibitem[Helm et~al\mbox{.}(2025)]%
        {helm2025token}
\bibfield{author}{\bibinfo{person}{Falko Helm}, \bibinfo{person}{Nico Daheim}, {and} \bibinfo{person}{Iryna Gurevych}.} \bibinfo{year}{2025}\natexlab{}.
\newblock \showarticletitle{Token Weighting for Long-Range Language Modeling}. In \bibinfo{booktitle}{\emph{Findings of the Association for Computational Linguistics: NAACL 2025}}. \bibinfo{pages}{1440--1459}.
\newblock


\bibitem[Huang et~al\mbox{.}(2025)]%
        {huang2025logrules}
\bibfield{author}{\bibinfo{person}{Xin Huang}, \bibinfo{person}{Ting Zhang}, {and} \bibinfo{person}{Wen Zhao}.} \bibinfo{year}{2025}\natexlab{}.
\newblock \showarticletitle{LogRules: Enhancing Log Analysis Capability of Large Language Models through Rules}. In \bibinfo{booktitle}{\emph{Findings of the Association for Computational Linguistics: NAACL 2025}}. \bibinfo{pages}{452--470}.
\newblock


\bibitem[Huo et~al\mbox{.}(2023)]%
        {huo2023semparser}
\bibfield{author}{\bibinfo{person}{Yintong Huo}, \bibinfo{person}{Yuxin Su}, \bibinfo{person}{Cheryl Lee}, {and} \bibinfo{person}{Michael~R Lyu}.} \bibinfo{year}{2023}\natexlab{}.
\newblock \showarticletitle{SemParser: A Semantic Parser for Log Analytics}. In \bibinfo{booktitle}{\emph{2023 IEEE/ACM 45th International Conference on Software Engineering (ICSE)}}. IEEE, \bibinfo{pages}{881--893}.
\newblock


\bibitem[Jeffries et~al\mbox{.}(2013)]%
        {jeffries2013processes}
\bibfield{author}{\bibinfo{person}{Robin Jeffries}, \bibinfo{person}{Althea~A Turner}, \bibinfo{person}{Peter~G Poison}, {and} \bibinfo{person}{Michael~E Atwood}.} \bibinfo{year}{2013}\natexlab{}.
\newblock \showarticletitle{The processes involved in designing software}.
\newblock In \bibinfo{booktitle}{\emph{Cognitive skills and their acquisition}}. \bibinfo{publisher}{Psychology Press}, \bibinfo{pages}{255--283}.
\newblock


\bibitem[Ji et~al\mbox{.}(2025)]%
        {ji2025superlog}
\bibfield{author}{\bibinfo{person}{Yuhe Ji}, \bibinfo{person}{Yilun Liu}, \bibinfo{person}{Feiyu Yao}, \bibinfo{person}{Minggui He}, \bibinfo{person}{Shimin Tao}, \bibinfo{person}{Xiaofeng Zhao}, \bibinfo{person}{Su Chang}, \bibinfo{person}{Xinhua Yang}, \bibinfo{person}{Weibin Meng}, \bibinfo{person}{Yuming Xie}, \bibinfo{person}{Boxing Chen}, \bibinfo{person}{Shenglin Zhang}, {and} \bibinfo{person}{Yongqian Sun}.} \bibinfo{year}{2025}\natexlab{}.
\newblock \showarticletitle{Adapting Large Language Models to Log Analysis with Interpretable Domain Knowledge}. In \bibinfo{booktitle}{\emph{Proceedings of the 34rd ACM International Conference on Information and Knowledge Management}}.
\newblock


\bibitem[Jiang et~al\mbox{.}(2024)]%
        {jiang2024lilac}
\bibfield{author}{\bibinfo{person}{Zhihan Jiang}, \bibinfo{person}{Jinyang Liu}, \bibinfo{person}{Zhuangbin Chen}, \bibinfo{person}{Yichen Li}, \bibinfo{person}{Junjie Huang}, \bibinfo{person}{Yintong Huo}, \bibinfo{person}{Pinjia He}, \bibinfo{person}{Jiazhen Gu}, {and} \bibinfo{person}{Michael~R Lyu}.} \bibinfo{year}{2024}\natexlab{}.
\newblock \showarticletitle{LILAC: Log parsing using LLMs with adaptive parsing cache}.
\newblock \bibinfo{journal}{\emph{Proceedings of the ACM on Software Engineering}} \bibinfo{volume}{1}, \bibinfo{number}{FSE} (\bibinfo{year}{2024}), \bibinfo{pages}{137--160}.
\newblock


\bibitem[Kent and Souppaya(2006)]%
        {kent2006nist}
\bibfield{author}{\bibinfo{person}{Karen Kent} {and} \bibinfo{person}{Murugiah Souppaya}.} \bibinfo{year}{2006}\natexlab{}.
\newblock \showarticletitle{NIST SP 800--92, Guide to computer security log management}.
\newblock \bibinfo{journal}{\emph{National Institute of Standards and Technology (NIST)}} (\bibinfo{year}{2006}).
\newblock


\bibitem[Kong et~al\mbox{.}(2023)]%
        {kong2023better}
\bibfield{author}{\bibinfo{person}{Aobo Kong}, \bibinfo{person}{Shiwan Zhao}, \bibinfo{person}{Hao Chen}, \bibinfo{person}{Qicheng Li}, \bibinfo{person}{Yong Qin}, \bibinfo{person}{Ruiqi Sun}, \bibinfo{person}{Xin Zhou}, \bibinfo{person}{Enzhi Wang}, {and} \bibinfo{person}{Xiaohang Dong}.} \bibinfo{year}{2023}\natexlab{}.
\newblock \showarticletitle{Better zero-shot reasoning with role-play prompting}.
\newblock \bibinfo{journal}{\emph{arXiv preprint arXiv:2308.07702}} (\bibinfo{year}{2023}).
\newblock


\bibitem[Kudo et~al\mbox{.}(2024)]%
        {kudo2024think}
\bibfield{author}{\bibinfo{person}{Keito Kudo}, \bibinfo{person}{Yoichi Aoki}, \bibinfo{person}{Tatsuki Kuribayashi}, \bibinfo{person}{Shusaku Sone}, \bibinfo{person}{Masaya Taniguchi}, \bibinfo{person}{Ana Brassard}, \bibinfo{person}{Keisuke Sakaguchi}, {and} \bibinfo{person}{Kentaro Inui}.} \bibinfo{year}{2024}\natexlab{}.
\newblock \showarticletitle{Think-to-talk or talk-to-think? when llms come up with an answer in multi-step reasoning}.
\newblock \bibinfo{journal}{\emph{arXiv preprint arXiv:2412.01113}} (\bibinfo{year}{2024}).
\newblock


\bibitem[Le and Zhang(2022)]%
        {le2022log}
\bibfield{author}{\bibinfo{person}{Van-Hoang Le} {and} \bibinfo{person}{Hongyu Zhang}.} \bibinfo{year}{2022}\natexlab{}.
\newblock \showarticletitle{Log-based Anomaly Detection with Deep Learning: How Far Are We?}. In \bibinfo{booktitle}{\emph{2022 IEEE/ACM International Conference on Software Engineering (ICSE)}}. IEEE, \bibinfo{pages}{1356--1367}.
\newblock


\bibitem[Le and Zhang(2024)]%
        {le2024prelog}
\bibfield{author}{\bibinfo{person}{Van-Hoang Le} {and} \bibinfo{person}{Hongyu Zhang}.} \bibinfo{year}{2024}\natexlab{}.
\newblock \showarticletitle{Prelog: A pre-trained model for log analytics}.
\newblock \bibinfo{journal}{\emph{Proceedings of the ACM on Management of Data}} \bibinfo{volume}{2}, \bibinfo{number}{3} (\bibinfo{year}{2024}), \bibinfo{pages}{1--28}.
\newblock


\bibitem[Levenshtein et~al\mbox{.}(1966)]%
        {levenshtein1966binary}
\bibfield{author}{\bibinfo{person}{Vladimir~I Levenshtein} {et~al\mbox{.}}} \bibinfo{year}{1966}\natexlab{}.
\newblock \showarticletitle{Binary codes capable of correcting deletions, insertions, and reversals}. In \bibinfo{booktitle}{\emph{Soviet physics doklady}}, Vol.~\bibinfo{volume}{10}. Soviet Union, \bibinfo{pages}{707--710}.
\newblock


\bibitem[Li et~al\mbox{.}(2025)]%
        {li2025s}
\bibfield{author}{\bibinfo{person}{Dacheng Li}, \bibinfo{person}{Shiyi Cao}, \bibinfo{person}{Chengkun Cao}, \bibinfo{person}{Xiuyu Li}, \bibinfo{person}{Shangyin Tan}, \bibinfo{person}{Kurt Keutzer}, \bibinfo{person}{Jiarong Xing}, \bibinfo{person}{Joseph~E Gonzalez}, {and} \bibinfo{person}{Ion Stoica}.} \bibinfo{year}{2025}\natexlab{}.
\newblock \showarticletitle{S*: Test time scaling for code generation}.
\newblock \bibinfo{journal}{\emph{arXiv preprint arXiv:2502.14382}} (\bibinfo{year}{2025}).
\newblock


\bibitem[Li et~al\mbox{.}(2023)]%
        {li2023did}
\bibfield{author}{\bibinfo{person}{Zhenhao Li}, \bibinfo{person}{Chuan Luo}, \bibinfo{person}{Tse-Hsun~Peter Chen}, \bibinfo{person}{Weiyi Shang}, \bibinfo{person}{Shilin He}, \bibinfo{person}{Qingwei Lin}, {and} \bibinfo{person}{Dongmei Zhang}.} \bibinfo{year}{2023}\natexlab{}.
\newblock \showarticletitle{Did We Miss Something Important? Studying and Exploring Variable-Aware Log Abstraction}. In \bibinfo{booktitle}{\emph{ICSE 2023}}.
\newblock


\bibitem[Lin(2004)]%
        {lin2004rouge}
\bibfield{author}{\bibinfo{person}{Chin-Yew Lin}.} \bibinfo{year}{2004}\natexlab{}.
\newblock \showarticletitle{Rouge: A package for automatic evaluation of summaries}. In \bibinfo{booktitle}{\emph{Text summarization branches out}}. \bibinfo{pages}{74--81}.
\newblock


\bibitem[Liu et~al\mbox{.}(2024a)]%
        {liu2024you}
\bibfield{author}{\bibinfo{person}{Yilun Liu}, \bibinfo{person}{Minggui He}, \bibinfo{person}{Feiyu Yao}, \bibinfo{person}{Yuhe Ji}, \bibinfo{person}{Shimin Tao}, \bibinfo{person}{Jingzhou Du}, \bibinfo{person}{Duan Li}, \bibinfo{person}{Jian Gao}, \bibinfo{person}{Li Zhang}, \bibinfo{person}{Hao Yang}, {et~al\mbox{.}}} \bibinfo{year}{2024}\natexlab{a}.
\newblock \showarticletitle{What Do You Want? User-centric Prompt Generation for Text-to-image Synthesis via Multi-turn Guidance}.
\newblock \bibinfo{journal}{\emph{arXiv preprint arXiv:2408.12910}} (\bibinfo{year}{2024}).
\newblock


\bibitem[Liu et~al\mbox{.}(2025b)]%
        {liu2025loglm}
\bibfield{author}{\bibinfo{person}{Yilun Liu}, \bibinfo{person}{Yuhe Ji}, \bibinfo{person}{Shimin Tao}, \bibinfo{person}{Minggui He}, \bibinfo{person}{Weibin Meng}, \bibinfo{person}{Shenglin Zhang}, \bibinfo{person}{Yongqian Sun}, \bibinfo{person}{Yuming Xie}, \bibinfo{person}{Boxing Chen}, {and} \bibinfo{person}{Hao Yang}.} \bibinfo{year}{2025}\natexlab{b}.
\newblock \showarticletitle{Loglm: From task-based to instruction-based automated log analysis}. In \bibinfo{booktitle}{\emph{2025 IEEE/ACM 47th International Conference on Software Engineering: Software Engineering in Practice (ICSE-SEIP)}}. IEEE, \bibinfo{pages}{401--412}.
\newblock


\bibitem[Liu et~al\mbox{.}(2024b)]%
        {liu2024interpretable}
\bibfield{author}{\bibinfo{person}{Yilun Liu}, \bibinfo{person}{Shimin Tao}, \bibinfo{person}{Weibin Meng}, \bibinfo{person}{Jingyu Wang}, \bibinfo{person}{Wenbing Ma}, \bibinfo{person}{Yuhang Chen}, \bibinfo{person}{Yanqing Zhao}, \bibinfo{person}{Hao Yang}, {and} \bibinfo{person}{Yanfei Jiang}.} \bibinfo{year}{2024}\natexlab{b}.
\newblock \showarticletitle{Interpretable online log analysis using large language models with prompt strategies}. In \bibinfo{booktitle}{\emph{Proceedings of the 32nd IEEE/ACM International Conference on Program Comprehension}}. \bibinfo{pages}{35--46}.
\newblock


\bibitem[Liu et~al\mbox{.}(2024c)]%
        {liu2024logprompt}
\bibfield{author}{\bibinfo{person}{Yilun Liu}, \bibinfo{person}{Shimin Tao}, \bibinfo{person}{Weibin Meng}, \bibinfo{person}{Feiyu Yao}, \bibinfo{person}{Xiaofeng Zhao}, {and} \bibinfo{person}{Hao Yang}.} \bibinfo{year}{2024}\natexlab{c}.
\newblock \showarticletitle{Logprompt: Prompt engineering towards zero-shot and interpretable log analysis}. In \bibinfo{booktitle}{\emph{Proceedings of the 2024 IEEE/ACM 46th International Conference on Software Engineering: Companion Proceedings}}. \bibinfo{pages}{364--365}.
\newblock


\bibitem[Liu et~al\mbox{.}(2025a)]%
        {liu2025fin}
\bibfield{author}{\bibinfo{person}{Zhaowei Liu}, \bibinfo{person}{Xin Guo}, \bibinfo{person}{Fangqi Lou}, \bibinfo{person}{Lingfeng Zeng}, \bibinfo{person}{Jinyi Niu}, \bibinfo{person}{Zixuan Wang}, \bibinfo{person}{Jiajie Xu}, \bibinfo{person}{Weige Cai}, \bibinfo{person}{Ziwei Yang}, \bibinfo{person}{Xueqian Zhao}, {et~al\mbox{.}}} \bibinfo{year}{2025}\natexlab{a}.
\newblock \showarticletitle{Fin-r1: A large language model for financial reasoning through reinforcement learning}.
\newblock \bibinfo{journal}{\emph{arXiv preprint arXiv:2503.16252}} (\bibinfo{year}{2025}).
\newblock


\bibitem[Ma et~al\mbox{.}(2025)]%
        {ma2025practitioners}
\bibfield{author}{\bibinfo{person}{Xiaoxue Ma}, \bibinfo{person}{Yishu Li}, \bibinfo{person}{Jacky Keung}, \bibinfo{person}{Xiao Yu}, \bibinfo{person}{Huiqi Zou}, \bibinfo{person}{Zhen Yang}, \bibinfo{person}{Federica Sarro}, {and} \bibinfo{person}{Earl~T Barr}.} \bibinfo{year}{2025}\natexlab{}.
\newblock \showarticletitle{Practitioners’ expectations on log anomaly detection}.
\newblock \bibinfo{journal}{\emph{IEEE Transactions on Software Engineering}} (\bibinfo{year}{2025}).
\newblock


\bibitem[Makanju et~al\mbox{.}(2009)]%
        {makanju2009clustering}
\bibfield{author}{\bibinfo{person}{Adetokunbo~AO Makanju}, \bibinfo{person}{A~Nur Zincir-Heywood}, {and} \bibinfo{person}{Evangelos~E Milios}.} \bibinfo{year}{2009}\natexlab{}.
\newblock \showarticletitle{Clustering event logs using iterative partitioning}. In \bibinfo{booktitle}{\emph{Proceedings of the 15th ACM SIGKDD international conference on Knowledge discovery and data mining}}. \bibinfo{pages}{1255--1264}.
\newblock


\bibitem[Meng et~al\mbox{.}(2020a)]%
        {meng2020logparse}
\bibfield{author}{\bibinfo{person}{Weibin Meng}, \bibinfo{person}{Ying Liu}, \bibinfo{person}{Federico Zaiter}, {et~al\mbox{.}}} \bibinfo{year}{2020}\natexlab{a}.
\newblock \showarticletitle{Logparse: Making log parsing adaptive through word classification}. In \bibinfo{booktitle}{\emph{2020 29th International Conference on Computer Communications and Networks (ICCCN)}}. \bibinfo{pages}{1--9}.
\newblock


\bibitem[Meng et~al\mbox{.}(2019)]%
        {meng2019loganomaly}
\bibfield{author}{\bibinfo{person}{Weibin Meng}, \bibinfo{person}{Ying Liu}, \bibinfo{person}{Yichen Zhu}, {et~al\mbox{.}}} \bibinfo{year}{2019}\natexlab{}.
\newblock \showarticletitle{LogAnomaly: Unsupervised detection of sequential and quantitative anomalies in unstructured logs.}. In \bibinfo{booktitle}{\emph{IJCAI}}, Vol.~\bibinfo{volume}{19}. \bibinfo{pages}{4739--4745}.
\newblock


\bibitem[Meng et~al\mbox{.}(2020b)]%
        {meng2020summarizing}
\bibfield{author}{\bibinfo{person}{Weibin Meng}, \bibinfo{person}{Federico Zaiter}, \bibinfo{person}{Yuheng Huang}, \bibinfo{person}{Ying Liu}, \bibinfo{person}{Shenglin Zhang}, \bibinfo{person}{Yuzhe Zhang}, \bibinfo{person}{Yichen Zhu}, \bibinfo{person}{Tianke Zhang}, \bibinfo{person}{En Wang}, \bibinfo{person}{Zuomin Ren}, {et~al\mbox{.}}} \bibinfo{year}{2020}\natexlab{b}.
\newblock \showarticletitle{Summarizing unstructured logs in online services}.
\newblock \bibinfo{journal}{\emph{arXiv preprint arXiv:2012.08938}} (\bibinfo{year}{2020}).
\newblock


\bibitem[Oliner and Stearley(2007)]%
        {oliner2007supercomputers}
\bibfield{author}{\bibinfo{person}{Adam Oliner} {and} \bibinfo{person}{Jon Stearley}.} \bibinfo{year}{2007}\natexlab{}.
\newblock \showarticletitle{What supercomputers say: A study of five system logs}. In \bibinfo{booktitle}{\emph{37th annual IEEE/IFIP international conference on dependable systems and networks (DSN'07)}}. IEEE, \bibinfo{pages}{575--584}.
\newblock


\bibitem[Pan et~al\mbox{.}(2024)]%
        {pan2024raglog}
\bibfield{author}{\bibinfo{person}{Jonathan Pan}, \bibinfo{person}{Wong~Swee Liang}, {and} \bibinfo{person}{Yuan Yidi}.} \bibinfo{year}{2024}\natexlab{}.
\newblock \showarticletitle{RAGLog: Log Anomaly Detection using Retrieval Augmented Generation}. In \bibinfo{booktitle}{\emph{2024 IEEE World Forum on Public Safety Technology (WFPST)}}. IEEE, \bibinfo{pages}{169--174}.
\newblock


\bibitem[Papineni et~al\mbox{.}(2002)]%
        {papineni2002bleu}
\bibfield{author}{\bibinfo{person}{Kishore Papineni}, \bibinfo{person}{Salim Roukos}, \bibinfo{person}{Todd Ward}, {and} \bibinfo{person}{Wei-Jing Zhu}.} \bibinfo{year}{2002}\natexlab{}.
\newblock \showarticletitle{Bleu: a method for automatic evaluation of machine translation}. In \bibinfo{booktitle}{\emph{Proceedings of the 40th annual meeting of the Association for Computational Linguistics}}. \bibinfo{pages}{311--318}.
\newblock


\bibitem[Qi et~al\mbox{.}(2023)]%
        {qi2023loggpt}
\bibfield{author}{\bibinfo{person}{Jiaxing Qi}, \bibinfo{person}{Shaohan Huang}, \bibinfo{person}{Zhongzhi Luan}, \bibinfo{person}{Shu Yang}, \bibinfo{person}{Carol Fung}, \bibinfo{person}{Hailong Yang}, \bibinfo{person}{Depei Qian}, \bibinfo{person}{Jing Shang}, \bibinfo{person}{Zhiwen Xiao}, {and} \bibinfo{person}{Zhihui Wu}.} \bibinfo{year}{2023}\natexlab{}.
\newblock \showarticletitle{Loggpt: Exploring chatgpt for log-based anomaly detection}. In \bibinfo{booktitle}{\emph{2023 IEEE International Conference on High Performance Computing \& Communications, Data Science \& Systems, Smart City \& Dependability in Sensor, Cloud \& Big Data Systems \& Application (HPCC/DSS/SmartCity/DependSys)}}. IEEE, \bibinfo{pages}{273--280}.
\newblock


\bibitem[Qian et~al\mbox{.}(2025)]%
        {qian2025fino1}
\bibfield{author}{\bibinfo{person}{Lingfei Qian}, \bibinfo{person}{Weipeng Zhou}, \bibinfo{person}{Yan Wang}, \bibinfo{person}{Xueqing Peng}, \bibinfo{person}{Jimin Huang}, {and} \bibinfo{person}{Qianqian Xie}.} \bibinfo{year}{2025}\natexlab{}.
\newblock \showarticletitle{Fino1: On the transferability of reasoning enhanced llms to finance}.
\newblock \bibinfo{journal}{\emph{arXiv e-prints}} (\bibinfo{year}{2025}), \bibinfo{pages}{arXiv--2502}.
\newblock


\bibitem[Rand(1971)]%
        {rand1971objective}
\bibfield{author}{\bibinfo{person}{William~M Rand}.} \bibinfo{year}{1971}\natexlab{}.
\newblock \showarticletitle{Objective criteria for the evaluation of clustering methods}.
\newblock \bibinfo{journal}{\emph{Journal of the American Statistical association}} \bibinfo{volume}{66}, \bibinfo{number}{336} (\bibinfo{year}{1971}), \bibinfo{pages}{846--850}.
\newblock


\bibitem[Schick et~al\mbox{.}(2024)]%
        {schick2024toolformer}
\bibfield{author}{\bibinfo{person}{Timo Schick}, \bibinfo{person}{Jane Dwivedi-Yu}, \bibinfo{person}{Roberto Dess{\`\i}}, \bibinfo{person}{Roberta Raileanu}, \bibinfo{person}{Maria Lomeli}, \bibinfo{person}{Eric Hambro}, \bibinfo{person}{Luke Zettlemoyer}, \bibinfo{person}{Nicola Cancedda}, {and} \bibinfo{person}{Thomas Scialom}.} \bibinfo{year}{2024}\natexlab{}.
\newblock \showarticletitle{Toolformer: Language models can teach themselves to use tools}.
\newblock \bibinfo{journal}{\emph{Advances in Neural Information Processing Systems}}  \bibinfo{volume}{36} (\bibinfo{year}{2024}).
\newblock


\bibitem[Shao et~al\mbox{.}(2024)]%
        {shao2024deepseekmath}
\bibfield{author}{\bibinfo{person}{Zhihong Shao}, \bibinfo{person}{Peiyi Wang}, \bibinfo{person}{Qihao Zhu}, \bibinfo{person}{Runxin Xu}, \bibinfo{person}{Junxiao Song}, \bibinfo{person}{Xiao Bi}, \bibinfo{person}{Haowei Zhang}, \bibinfo{person}{Mingchuan Zhang}, \bibinfo{person}{YK Li}, \bibinfo{person}{Yang Wu}, {et~al\mbox{.}}} \bibinfo{year}{2024}\natexlab{}.
\newblock \showarticletitle{Deepseekmath: Pushing the limits of mathematical reasoning in open language models}.
\newblock \bibinfo{journal}{\emph{arXiv preprint arXiv:2402.03300}} (\bibinfo{year}{2024}).
\newblock


\bibitem[Sheng et~al\mbox{.}(2024)]%
        {sheng2024hybridflow}
\bibfield{author}{\bibinfo{person}{Guangming Sheng}, \bibinfo{person}{Chi Zhang}, \bibinfo{person}{Zilingfeng Ye}, \bibinfo{person}{Xibin Wu}, \bibinfo{person}{Wang Zhang}, \bibinfo{person}{Ru Zhang}, \bibinfo{person}{Yanghua Peng}, \bibinfo{person}{Haibin Lin}, {and} \bibinfo{person}{Chuan Wu}.} \bibinfo{year}{2024}\natexlab{}.
\newblock \showarticletitle{HybridFlow: A Flexible and Efficient RLHF Framework}.
\newblock \bibinfo{journal}{\emph{arXiv preprint arXiv: 2409.19256}} (\bibinfo{year}{2024}).
\newblock


\bibitem[Smith et~al\mbox{.}(2024)]%
        {smithunderstanding}
\bibfield{author}{\bibinfo{person}{Alexander Smith}, \bibinfo{person}{William Martinez}, \bibinfo{person}{Sophia Garcia}, \bibinfo{person}{Benjamin Thomas}, \bibinfo{person}{Olivia Davis}, {and} \bibinfo{person}{Weimang Ye}.} \bibinfo{year}{2024}\natexlab{}.
\newblock \showarticletitle{Understanding Distribution Shift in LLMs: Methods, Evaluations, and Challenges}.
\newblock \bibinfo{journal}{\emph{Preprint on ResearchGate}} (\bibinfo{date}{02} \bibinfo{year}{2024}).
\newblock
\href{https://doi.org/10.13140/RG.2.2.33962.32963}{doi:\nolinkurl{10.13140/RG.2.2.33962.32963}}


\bibitem[Sun et~al\mbox{.}(2025)]%
        {sun2025enhancing}
\bibfield{author}{\bibinfo{person}{Yongqian Sun}, \bibinfo{person}{Weihua Kuang}, \bibinfo{person}{Chao Shen}, \bibinfo{person}{Xidao Wen}, \bibinfo{person}{Tinghua Zheng}, \bibinfo{person}{Heng Liu}, \bibinfo{person}{Shenglin Zhang}, \bibinfo{person}{Bo Wu}, {and} \bibinfo{person}{Dan Pei}.} \bibinfo{year}{2025}\natexlab{}.
\newblock \showarticletitle{Enhancing Interpretability in Software Change Management with Chain-of-Thought Reasoning}.
\newblock \bibinfo{journal}{\emph{arXiv preprint arXiv:2507.09315}} (\bibinfo{year}{2025}).
\newblock


\bibitem[Tall et~al\mbox{.}(2001)]%
        {tall2001symbols}
\bibfield{author}{\bibinfo{person}{David Tall}, \bibinfo{person}{Eddie Gray}, \bibinfo{person}{Maselan~Bin Ali}, \bibinfo{person}{Lillie Crowley}, \bibinfo{person}{Phil DeMarois}, \bibinfo{person}{Mercedes McGowen}, \bibinfo{person}{Demetra Pitta}, \bibinfo{person}{Marcia Pinto}, \bibinfo{person}{Michael Thomas}, {and} \bibinfo{person}{Yudariah Yusof}.} \bibinfo{year}{2001}\natexlab{}.
\newblock \showarticletitle{Symbols and the bifurcation between procedural and conceptual thinking}.
\newblock \bibinfo{journal}{\emph{Canadian Journal of Math, Science \& Technology Education}} \bibinfo{volume}{1}, \bibinfo{number}{1} (\bibinfo{year}{2001}), \bibinfo{pages}{81--104}.
\newblock


\bibitem[Tao et~al\mbox{.}(2023)]%
        {tao2023biglog}
\bibfield{author}{\bibinfo{person}{Shimin Tao}, \bibinfo{person}{Yilun Liu}, \bibinfo{person}{Weibin Meng}, \bibinfo{person}{Zuomin Ren}, \bibinfo{person}{Hao Yang}, \bibinfo{person}{Xun Chen}, \bibinfo{person}{Liang Zhang}, \bibinfo{person}{Yuming Xie}, \bibinfo{person}{Chang Su}, \bibinfo{person}{Xiaosong Oiao}, {et~al\mbox{.}}} \bibinfo{year}{2023}\natexlab{}.
\newblock \showarticletitle{Biglog: Unsupervised large-scale pre-training for a unified log representation}. In \bibinfo{booktitle}{\emph{2023 IEEE/ACM 31st International Symposium on Quality of Service (IWQoS)}}. IEEE, \bibinfo{pages}{1--11}.
\newblock


\bibitem[Tao et~al\mbox{.}(2022)]%
        {tao2022logstamp}
\bibfield{author}{\bibinfo{person}{Shimin Tao}, \bibinfo{person}{Weibin Meng}, \bibinfo{person}{Yimeng Cheng}, \bibinfo{person}{Yichen Zhu}, \bibinfo{person}{Ying Liu}, \bibinfo{person}{Chunning Du}, \bibinfo{person}{Tao Han}, \bibinfo{person}{Yongpeng Zhao}, \bibinfo{person}{Xiangguang Wang}, {and} \bibinfo{person}{Hao Yang}.} \bibinfo{year}{2022}\natexlab{}.
\newblock \showarticletitle{Logstamp: Automatic online log parsing based on sequence labelling}.
\newblock \bibinfo{journal}{\emph{ACM SIGMETRICS Performance Evaluation Review}} \bibinfo{volume}{49}, \bibinfo{number}{4} (\bibinfo{year}{2022}), \bibinfo{pages}{93--98}.
\newblock


\bibitem[Team(2025)]%
        {qwen3technicalreport}
\bibfield{author}{\bibinfo{person}{Qwen Team}.} \bibinfo{year}{2025}\natexlab{}.
\newblock \bibinfo{title}{Qwen3 Technical Report}.
\newblock
\showeprint[arxiv]{2505.09388}~[cs.CL]
\urldef\tempurl%
\url{https://arxiv.org/abs/2505.09388}
\showURL{%
\tempurl}


\bibitem[van Linschoten(2024)]%
        {The_State_of_LLM_Operations_or_LLMOps}
\bibfield{author}{\bibinfo{person}{Alex~Strick van Linschoten}.} \bibinfo{year}{2024}\natexlab{}.
\newblock \bibinfo{title}{The State of LLM Operations or LLMOps: Why Everything is Hard}.
\newblock \bibinfo{howpublished}{https://www.zenml.io/blog/state-of-llmops-why-everything-is-hard}.
\newblock


\bibitem[Wang et~al\mbox{.}(2020)]%
        {wang2020overview}
\bibfield{author}{\bibinfo{person}{Haoran Wang}, \bibinfo{person}{Yue Zhang}, {and} \bibinfo{person}{Xiaosheng Yu}.} \bibinfo{year}{2020}\natexlab{}.
\newblock \showarticletitle{An overview of image caption generation methods}.
\newblock \bibinfo{journal}{\emph{Computational intelligence and neuroscience}} \bibinfo{volume}{2020}, \bibinfo{number}{1} (\bibinfo{year}{2020}), \bibinfo{pages}{3062706}.
\newblock


\bibitem[Wang et~al\mbox{.}(2024)]%
        {wang2024logexpert}
\bibfield{author}{\bibinfo{person}{Jiabo Wang}, \bibinfo{person}{Guojun Chu}, \bibinfo{person}{Jingyu Wang}, \bibinfo{person}{Haifeng Sun}, \bibinfo{person}{Qi Qi}, \bibinfo{person}{Yuanyi Wang}, \bibinfo{person}{Ji Qi}, {and} \bibinfo{person}{Jianxin Liao}.} \bibinfo{year}{2024}\natexlab{}.
\newblock \showarticletitle{LogExpert: Log-based Recommended Resolutions Generation using Large Language Model}. In \bibinfo{booktitle}{\emph{Proceedings of the 2024 ACM/IEEE 44th International Conference on Software Engineering: New Ideas and Emerging Results}}. \bibinfo{pages}{42--46}.
\newblock


\bibitem[Wang et~al\mbox{.}(2022)]%
        {wang2022deep}
\bibfield{author}{\bibinfo{person}{Xu Wang}, \bibinfo{person}{Sen Wang}, \bibinfo{person}{Xingxing Liang}, \bibinfo{person}{Dawei Zhao}, \bibinfo{person}{Jincai Huang}, \bibinfo{person}{Xin Xu}, \bibinfo{person}{Bin Dai}, {and} \bibinfo{person}{Qiguang Miao}.} \bibinfo{year}{2022}\natexlab{}.
\newblock \showarticletitle{Deep reinforcement learning: A survey}.
\newblock \bibinfo{journal}{\emph{IEEE Transactions on Neural Networks and Learning Systems}} \bibinfo{volume}{35}, \bibinfo{number}{4} (\bibinfo{year}{2022}), \bibinfo{pages}{5064--5078}.
\newblock


\bibitem[Xu et~al\mbox{.}(2025)]%
        {xu2025rationanomalyloganomalydetection}
\bibfield{author}{\bibinfo{person}{Song Xu}, \bibinfo{person}{Yilun Liu}, \bibinfo{person}{Minggui He}, \bibinfo{person}{Mingchen Dai}, \bibinfo{person}{Ziang Chen}, \bibinfo{person}{Chunguang Zhao}, \bibinfo{person}{Jingzhou Du}, \bibinfo{person}{Shimin Tao}, \bibinfo{person}{Weibin Meng}, \bibinfo{person}{Shenglin Zhang}, \bibinfo{person}{Yongqian Sun}, \bibinfo{person}{Boxing Chen}, {and} \bibinfo{person}{Daimeng Wei}.} \bibinfo{year}{2025}\natexlab{}.
\newblock \bibinfo{title}{RationAnomaly: Log Anomaly Detection with Rationality via Chain-of-Thought and Reinforcement Learning}.
\newblock


\bibitem[Yang et~al\mbox{.}(2024)]%
        {Yang2024Qwen25TR}
\bibfield{author}{\bibinfo{person}{An Yang}, \bibinfo{person}{Baosong Yang}, \bibinfo{person}{Beichen Zhang}, {and} \bibinfo{person}{et~al. Binyuan~Hui}.} \bibinfo{year}{2024}\natexlab{}.
\newblock \showarticletitle{Qwen2.5 Technical Report}.
\newblock \bibinfo{journal}{\emph{ArXiv}}  \bibinfo{volume}{abs/2412.15115} (\bibinfo{year}{2024}).
\newblock
\urldef\tempurl%
\url{https://api.semanticscholar.org/CorpusID:274859421}
\showURL{%
\tempurl}


\bibitem[Zhang et~al\mbox{.}(2025)]%
        {zhang2025thinkfl}
\bibfield{author}{\bibinfo{person}{Lingzhe Zhang}, \bibinfo{person}{Yunpeng Zhai}, \bibinfo{person}{Tong Jia}, \bibinfo{person}{Chiming Duan}, \bibinfo{person}{Siyu Yu}, \bibinfo{person}{Jinyang Gao}, \bibinfo{person}{Bolin Ding}, \bibinfo{person}{Zhonghai Wu}, {and} \bibinfo{person}{Ying Li}.} \bibinfo{year}{2025}\natexlab{}.
\newblock \showarticletitle{ThinkFL: Self-Refining Failure Localization for Microservice Systems via Reinforcement Fine-Tuning}.
\newblock \bibinfo{journal}{\emph{arXiv preprint arXiv:2504.18776}} (\bibinfo{year}{2025}).
\newblock


\bibitem[Zhang et~al\mbox{.}(2007)]%
        {zhang2017syslog}
\bibfield{author}{\bibinfo{person}{Shenglin Zhang}, \bibinfo{person}{Weibin Meng}, {et~al\mbox{.}}} \bibinfo{year}{2007}\natexlab{}.
\newblock \showarticletitle{Syslog processing for switch failure diagnosis and prediction in datacenter networks}. In \bibinfo{booktitle}{\emph{IEEE/ACM 25th International Symposium on Quality of Service (IWQoS'17)}}. \bibinfo{pages}{1--10}.
\newblock


\bibitem[Zhang et~al\mbox{.}(2019)]%
        {zhang2019robust}
\bibfield{author}{\bibinfo{person}{Xu Zhang}, \bibinfo{person}{Yong Xu}, \bibinfo{person}{Qingwei Lin}, \bibinfo{person}{Bo Qiao}, \bibinfo{person}{Hongyu Zhang}, \bibinfo{person}{Yingnong Dang}, \bibinfo{person}{Chunyu Xie}, \bibinfo{person}{Xinsheng Yang}, \bibinfo{person}{Qian Cheng}, \bibinfo{person}{Ze Li}, {et~al\mbox{.}}} \bibinfo{year}{2019}\natexlab{}.
\newblock \showarticletitle{Robust log-based anomaly detection on unstable log data}. In \bibinfo{booktitle}{\emph{Proceedings of the 2019 27th ACM Joint Meeting on European Software Engineering Conference and Symposium on the Foundations of Software Engineering}}. \bibinfo{pages}{807--817}.
\newblock


\bibitem[Zheng et~al\mbox{.}(2024)]%
        {zheng-etal-2024-llamafactory}
\bibfield{author}{\bibinfo{person}{Yaowei Zheng}, \bibinfo{person}{Richong Zhang}, \bibinfo{person}{Junhao Zhang}, \bibinfo{person}{YeYanhan YeYanhan}, {and} \bibinfo{person}{Zheyan Luo}.} \bibinfo{year}{2024}\natexlab{}.
\newblock \showarticletitle{{L}lama{F}actory: Unified Efficient Fine-Tuning of 100+ Language Models}. In \bibinfo{booktitle}{\emph{Proceedings of the 62nd Annual Meeting of the Association for Computational Linguistics (Volume 3: System Demonstrations)}}, \bibfield{editor}{\bibinfo{person}{Yixin Cao}, \bibinfo{person}{Yang Feng}, {and} \bibinfo{person}{Deyi Xiong}} (Eds.). \bibinfo{publisher}{Association for Computational Linguistics}, \bibinfo{address}{Bangkok, Thailand}, \bibinfo{pages}{400--410}.
\newblock
\href{https://doi.org/10.18653/v1/2024.acl-demos.38}{doi:\nolinkurl{10.18653/v1/2024.acl-demos.38}}


\bibitem[Zhu et~al\mbox{.}(2019)]%
        {zhu2019tools}
\bibfield{author}{\bibinfo{person}{Jieming Zhu}, \bibinfo{person}{Shilin He}, \bibinfo{person}{Jinyang Liu}, \bibinfo{person}{Pinjia He}, \bibinfo{person}{Qi Xie}, \bibinfo{person}{Zibin Zheng}, {and} \bibinfo{person}{Michael~R Lyu}.} \bibinfo{year}{2019}\natexlab{}.
\newblock \showarticletitle{Tools and benchmarks for automated log parsing}. In \bibinfo{booktitle}{\emph{2019 IEEE/ACM 41st International Conference on Software Engineering: Software Engineering in Practice (ICSE-SEIP)}}. IEEE, \bibinfo{pages}{121--130}.
\newblock


\bibitem[Zhu et~al\mbox{.}(2021)]%
        {zhu2021unilog}
\bibfield{author}{\bibinfo{person}{Yichen Zhu}, \bibinfo{person}{Weibin Meng}, \bibinfo{person}{Ying Liu}, \bibinfo{person}{Shenglin Zhang}, \bibinfo{person}{Tao Han}, \bibinfo{person}{Shimin Tao}, {and} \bibinfo{person}{Dan Pei}.} \bibinfo{year}{2021}\natexlab{}.
\newblock \showarticletitle{Unilog: Deploy one model and specialize it for all log analysis tasks}.
\newblock \bibinfo{journal}{\emph{arXiv preprint arXiv:2112.03159}} (\bibinfo{year}{2021}).
\newblock


\end{thebibliography}

\end{document}